%% file: bare_conf_compsoc.tex
\documentclass[conference,compsoc]{IEEEtran}
\ifCLASSOPTIONcompsoc
  \usepackage[nocompress]{cite}
\else
  \usepackage{cite}
\fi
\ifCLASSINFOpdf
\else
\fi

\usepackage{cite}
\usepackage[normalem]{ulem}
\useunder{\uline}{\ul}{}
\usepackage{amsmath,amssymb,amsfonts}
\usepackage{algorithm}
\usepackage{algpseudocode}
\usepackage{graphicx}
\usepackage{textcomp}
\usepackage{bmpsize}
\usepackage{xcolor}
\usepackage{lipsum}
\usepackage{amsthm}
\usepackage{caption}
\usepackage{bbm}
\usepackage{wasysym}
\usepackage{booktabs}
\usepackage{multirow}
\usepackage{enumitem}
\usepackage{subcaption}
\usepackage{arydshln}
\usepackage{tabularray}
\usepackage[numbers]{natbib}
\usepackage[most]{tcolorbox}

\usepackage[colorlinks=true,urlcolor=black]{hyperref}
\def\BibTeX{{\rm B\kern-.05em{\sc i\kern-.025em b}\kern-.08em
    T\kern-.1667em\lower.7ex\hbox{E}\kern-.125emX}}

\begin{document}
%

\title{Privacy Vulnerabilities of Attention Layers in Tabular Foundation Models and Protection of High-Risk Queries}




\author{\IEEEauthorblockN{Tânia Carvalho}
\IEEEauthorblockA{SnT, University of Luxembourg \\ Luxembourg \\
}
\and
\IEEEauthorblockN{Maxime Cordy}
\IEEEauthorblockA{SnT, University of Luxembourg \\ Luxembourg\\
}
}


%


\maketitle

\begin{abstract}
Tabular foundation models are commonly assumed to present limited privacy concerns as they are often pre-trained on large collections of synthetic data. However, these models leverage in-context learning, where sensitive records may be provided directly at inference time as labelled context examples. In this paper, we demonstrate that predictions generated via the attention mechanism leak sufficient information to enable effective Membership Inference Attacks (MIAs). To highlight this vulnerability, we propose AMIA (Attention-based Membership Inference Attack), a shadow-model-free attack that exploits the concentration of transformer attention patterns. Our results show that attention mechanisms reveal strong membership signals, which exceed classical confidence-based attacks, achieving an average gain of 7.7\%, specially in low false-positive regimes.
To mitigate this risk, we introduce an inference-time defence inspired by $k$-anonymity principles. This approach reduces the uniqueness of context-key representations without introducing random noise or retraining the model. By targeting only high-risk queries identified through AMIA scores, the defence substantially reduces membership leakage of this attack by an average of 50\% and 25\% against confidence-based attacks, while preserving predictive utility with only 3.9\% performance degradation.
Beyond showing that context examples are vulnerable, we further demonstrate that fine-tuning introduces an additional source of privacy risk. In particular, samples whose prediction confidence increases after fine-tuning become more susceptible to MIAs, indicating that fine-tuning can amplify memorisation and expose sensitive training information through confidence shifts.

\end{abstract}



%
\IEEEpeerreviewmaketitle

\section{Introduction}
Tabular data is the predominant data format in many machine learning applications, particularly in domains where records correspond directly to individuals, i.e. microdata. It plays a central role in privacy research, as sensitive and socially critical sectors such as healthcare, financial transactions, census, education, and human resources rely heavily on tabular datasets for analysis and prediction. 

Unlike text or images, tabular data contains highly structured attributes, including quasi-identifiers, categorical variables and rare subgroups. These characteristics present unique challenges to machine learning models, such as heterogeneous feature types, missing values and skewed distributions. Consequently, tabular data can be particularly susceptible to privacy risks, including re-identification~\cite{carvalho2023survey}, attribute inference, and membership disclosure~\cite{hu2022membership}.

Recently, researchers have developed tabular foundation models (FMs), a new generation of models specifically designed for structured data. Building on the success of FMs in natural language processing, tabular FMs aim to learn transferable representations from diverse tabular datasets that can be applied to a variety of downstream tasks. Such models have demonstrated competitive performance on a wide range of classification benchmarks~\cite{erickson2026tabarena}, mainly due to their strong generalisation capabilities, which are attributed to large-scale pre-training and attention inference time mechanisms.

In contrast to many large language models (LLMs), tabular FMs are typically pre-trained on synthetic datasets rather than on real-world data. Besides contributing to high predictive performance, this design also favours privacy considerations. The generation of synthetic data aims to approximate real-world distributions while reducing the exposure of sensitive individual records.

However, this assumption can be misleading, given that synthetic pre-training shapes the model's weights but offers no protection for the real records provided as in-context examples.
In the tabular in-context learning (ICL) setting, the model attends over context examples to produce predictions for a query instance. Hence, privacy leakage does not necessarily arise from the model memorisation of the pre-training data, but rather from its dependence on context representations at inference time. 
This shifts the attack surface of membership inference attacks (MIAs) from parameter memorisation to context information leakage. 

We have conducted experiments on six diverse datasets, four classical machine learning models and four tabular FMs, under three standard MIAs.
Table~\ref{tab:rmia-lira-population-locations-dropout} provides initial evidence of this phenomenon by demonstrating that tabular FMs are vulnerable to existing standard MIAs.


    

Motivated by this observation, we move beyond conventional output-based MIAs and investigate whether attention patterns can reveal membership information more explicitly. In particular, we hypothesise that member queries induce disproportionately concentrated attention towards specific context records compared to non-members. Based on this intuition, we propose AMIA, an attention-based MIA that leverages attention dynamics across layers and heads to infer whether a queried example belongs to the private context set.

Our results show that AMIA consistently outperforms state-of-the-art MIAs while requiring less adversarial background knowledge and lower computational resources. AMIA is also robust to different context sizes, showing a stable AUC across different settings, whereas other attacks tend to degrade as the context increases.

To mitigate the identified leakage, we further propose a defence mechanism tailored to tabular ICL. 
Our approach applies label-aware microaggregation to $k$-anonymise context representations, but only when necessary. An adaptive high-risk guardrail uses AMIA scores to detect queries with concentrated attention and applies anonymisation exclusively to those with a score exceeding a threshold calibrated on context members, leaving low-risk queries unchanged. 


Our proposed target label $k$-anonymity defence effectively mitigates both AMIA and confidence-based attacks while having a negligible impact on predictive performance. The accuracy remains almost identical across models, indicating that the guardrail avoids the unnecessary anonymisation of innocuous queries. In the strongest settings, the defence reduces AMIA AUC close to zero with no accuracy loss, showing that high-risk context representations can be protected without degrading predictive performance.

While we evaluate MIAs against tabular FMs in their standard ICL deployment, practitioners may fine-tune models on sensitive data to improve task performance.
Although the objective is to adapt to specific data distributions, we demonstrate that fine-tuning can also introduce additional privacy risks. In particular, MIAs remain feasible after fine-tuning, as certain training samples induce disproportionately higher confidence scores and stronger predictive certainty. This behaviour increases the distinguishability between members and non-members, thereby amplifying the risk of membership leakage and exposing sensitive information about the fine-tuning dataset.

In summary, our contributions are as follows.

\begin{figure*}[!ht]
\centerline{\includegraphics[width=0.9\textwidth]{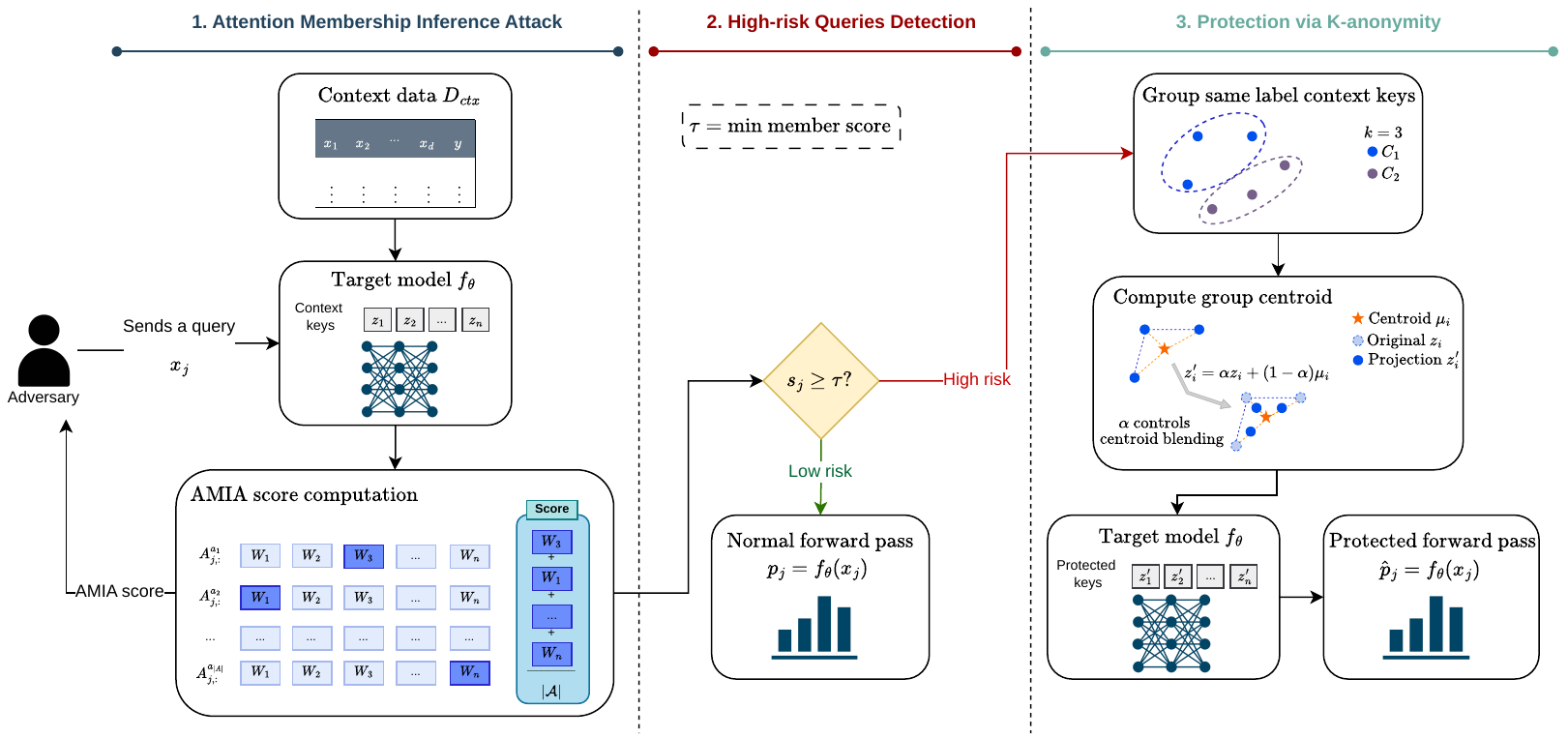}}
\caption{Overview of the proposed attention membership inference attack (AMIA) and defence for the high-risk queries. Phase 1 extracts the attention matrix over layers and heads, and then  the maximum attention weight is averaged across all (layer, head) pairs. Phase 2 selects the high-risk queries determined based on a threshold. Phase 3 applies a label-aware microaggretion to $k$-anonymise context keys.}
\label{fig:amia}
\end{figure*}

\begin{itemize}
    \item We provide the \textbf{first systematic evaluation of MIAs against tabular FMs} in the ICL setting, demonstrating that existing state-of-the-art MIAs remain effective despite synthetic pre-training.
    \item We \textbf{propose AMIA, a new attention-based MIA} that exploits attention dynamics across layers and heads to infer membership information from context set. Rather than relying on shadow models or auxiliary population data, the attack requires only access to the model.
    \item We design a \textbf{new inference-time} defence mechanism that adapts the principles of $k$-anonymity through label-aware microaggregation on high-risk queries (see Figure~\ref{fig:amia} for an overview of both AMIA and our defence). We evaluate our defence and show that it mitigates both confidence- based and attention-based attacks.
    \item We analyse the \textbf{privacy implications of fine-tuning tabular FMs} and demonstrate that it can amplify membership leakage by increasing confidence separation between specific member and non-member samples.

\end{itemize}


To foster reproducibility and transparency, the source code and datasets required to replicate the experiments presented in this paper are publicly available at https://github.com/serval-uni-lu/MIAonTabFMs.

\section{Preliminaries}

\subsection{In-context Learning} 
In-Context Learning (ICL) is an efficient paradigm for task-specific adaptation and is a distinctive capability of Large Language Models (LLMs)~\cite{brown2020language}. It enables models to perform new tasks using a small set of labelled examples, known as few-shot examples. In practice, ICL is an adaptation mechanism that operates at inference time, whereby a pre-trained model solves a downstream task using only the information provided in the input context. Unlike conventional supervised learning approaches, ICL does not require parameter updates or gradient-based optimisation. Instead, the model implicitly ``learns'' the task through attention over the contextual examples.
ICL has been studied in supervised learning settings, including regression~\cite{garg2022can,han2025understanding} and classification~\cite{reddy2023mechanistic,d2024context} tasks. 

In the classification setting, the input consists of a sequence of labelled context examples $D_{ctx} = (x_1, y_1), (x_2, y_2), ..., (x_L, y_L)$ followed by a query point $x_q$. Each $x_i \in \mathcal{X}$ belongs to an input space (e.g. text, images or tabular rows) and each label $y_i \in \mathcal{Y} = {1, \dots, C}$ corresponds to one of $C$ possible classes. Given the context set $D_{\text{ctx}}$ and the query sample $x_q$, the model predicts a probability distribution over the label space $P(y_q | x_q, D_{ctx})$. The predicted class is then obtained from the output logits over the $C$ candidate classes. The central intuition behind ICL is that the context implicitly specifies both the task structure and the decision rule required to solve the query.

Although LLMs can process tabular data within an ICL setting, several challenges make this process difficult in practice, including heterogeneous feature types, missing values and the lack of a natural sequential structure. To address these limitations, recent tabular foundation models (tabular FMs) have been specifically designed to perform ICL directly on structured tabular data. While tabular LLMs adapt language models to handle tables, tabular FMs are designed with tabular-specific transformers.

Models such as TabPFN~\cite{hollmann_tabpfn_2023, hollmann_tabpfn_2025} and TabICL~\cite{qu_tabicl_2025} are transformer-based architectures pre-trained to perform classification and regression through ICL on tabular tasks. In this setting, the training set acts as the context, while each test sample corresponds to a query instance. Formally, each training example is represented as a feature vector $x_i \in \mathbb{R}^d$, corresponding to the $i$-th record with $d$ features. These features may include both numerical and categorical attributes. Numerical features are typically normalised, while categorical variables are ordinal-encoded or embedded into continuous representations. Associated labels $y_i$ are incorporated through dedicated label embeddings. 

From a user perspective, tabular FMs expose interfaces similar to those of classical machine learning libraries, typically through \textit{fit()} and \textit{predict()} functions. 
This capability is enabled by transformer-based attention mechanisms that operate jointly across rows and features.
In particular, TabPFN uses a two-way attention mechanism whereby each cell attends both to the other features within the same row and to the corresponding feature across different rows~\cite{hollmann_tabpfn_2023}. 
The defining characteristic of tabular FMs is that they are meta-trained over a large collection of synthetic and real tabular tasks. As a result, the model learns a general inference procedure that can be applied to unseen datasets purely through in-context conditioning. Cross-row attention over the contextual training examples is therefore often sufficient for the model to infer an effective decision boundary.

Tabular FMs have demonstrated strong predictive performance, particularly in small data scenarios, where their ICL capabilities often outperform classical machine learning approaches~\cite{erickson2026tabarena}. This success has led to the development of several related architectures, including TabDPT~\cite{ma2026tabdpt}, TabSTAR~\cite{arazi2026tabstar}, Mitra~\cite{zhang2026mitra} and TabFlex~\cite{zeng2025tabflex}, among others.

\subsection{Membership Inference Attack}
MIAs are widely regarded as a standard framework for assessing privacy risks in machine learning systems~\cite{hu2022membership,hu2023defenses}. 
Given a trained target model \(f_\theta\), an adversary aims to determine whether a candidate record \(x_i\) was part of the target model's training dataset. Let $m_i \in \{0,1\}$ be the unknown ground-truth membership label of \(x_i\), where $m_i=1$ indicates that $x_i \in D_{\text{train}}$, and $m_i=0$ otherwise.

The attack objective is to construct a decision rule:
\[\mathcal{A}(x_i, f_\theta) \rightarrow \{0,1\}\]

As the true membership label \(m_i\) is unavailable to the adversary, the attack instead relies on observing the behaviour of the target model on \(x_i\). Specifically, the adversary computes a membership score:
\[s_i = \Lambda(x_i, f_\theta, \mathcal{K}),\]where \(\Lambda\) is a scoring function used by a specific attack, and \(\mathcal{K}\) denotes any auxiliary knowledge available to the adversary, such as reference models~\cite{carlini2022membership,zarifzadeh2023low}, population samples~\cite{ye2022enhanced}, or shadow data~\cite{shokri2017membership}.
The score \(s_i\) is evidence of membership. For instance, members may exhibit lower loss, higher confidence, or a larger likelihood ratio compared to non-members. The final attack prediction is then obtained by thresholding the score.


Several studies have demonstrated the vulnerability of ICL to MIAs, particularly in text classification settings~\cite{duan2024privacy, wen_membership_2024, chang2025context, ran2025lora}. In these attacks, the adversary aims to infer whether specific examples were included in the in-context demonstrations ($D_{ctx}$) by querying the model with candidate text sequences. 
Recent work has also investigated worst-case information leakage through adversarial canaries~\cite{choi2025contextleak}. In this setting, uniquely identifiable canary sequences are inserted into the context, and carefully designed adversarial queries are used to induce leakage of these sequences. The adversary can then infer the membership of the canary based on the model's responses.

Additionally, MIAs have been evaluated on LLMs fine-tuned on tabular data~\cite{German2025TabMIAAB}. One of the main challenges is the representation of the structured tables in a text-based input format suitable for transformer architectures. 

Despite the rapid adoption of tabular ICL, its privacy implications remain largely unexplored. In particular, no privacy evaluations currently target membership inference in this learning paradigm. Moreover, the application of MIAs in tabular ICL differs fundamentally from the text-based setting. Existing attacks against language models are typically evaluated through generation behaviour, where membership signals are extracted from token-level probabilities or sequence likelihoods~\cite{wen_membership_2024}. 

Both tabular- and text-based ICL models operate by keeping the pre-trained model parameters $\theta$ fixed 
and pass examples as context at inference. The difference is not the learning paradigm but the structure of the context entries and what that structure enables for membership inference. In text-ICL, each demonstration consists of a variable-length token sequence, where a single example may span multiple key-value positions within the attention mechanism.
On the other hand, tabular ICL operates on structured feature-label pairs $(x_j, y_j) \in \mathbb{R}^d \times \mathcal{Y}$ where each context example corresponds to a fixed-dimensional row. Consequently, it occupies exactly one key
position $K_j$ and one value position $V_j$.    
This raises the possibility that contextual tabular examples may exert a stronger and more detectable influence on model predictions, thereby increasing susceptibility to MIAs.

\subsection{Defences against MIA in ICL}
When privacy-preserving mechanisms are proposed to mitigate MIAs in ICL systems, Differential Privacy (DP) is generally considered as the gold standard due to its formal privacy guarantees~\cite{dwork2025differential}. DP provides a worst-case privacy analysis by bounding how much the output of a mechanism can change when a single individual's data is modified. In DP-based prediction settings, noise is introduced into the model's output distribution to ensure that the generated outputs satisfy DP. This framework is parameterised by the privacy budget \(\epsilon \in [0,\infty)\), where smaller values of \(\epsilon\) correspond to stronger privacy guarantees.

Several DP-based approaches for ICL have been proposed. The first DP-ICL~\cite{wu2024privacy} partitions the sensitive dataset into a collection of exemplars and produces differentially private outputs by aggregating the responses generated by the LLM over those examples. Other approaches locally protect labels through randomised response mechanisms~\cite{zheng2024locally}. More recent research has increasingly focused on generating privacy-preserving in-context demonstrations through synthetic data generation~\cite{amin2024private,bhusal2026privacy,tang2024privacy,Koskela2025DifferentiallyPI}. In parallel, prompt distillation construct private demonstrations by labeling publicly available datasets under DP constraints and using the newly labeled examples as in-context exemplars~\cite{duan2023flocks}. Nonetheless, all these DP-ICL approaches have been developed and evaluated in text-based settings. 
Furthermore, it has been shown that even when DP is incorporated into ICL systems, residual membership signals may still leak through model outputs and context interactions~\cite{choi2025contextleak,tang2024privacy}.

Beyond DP, recent work has also explored text-based ICL approaches to reduce leakage by constructing ensembles of independently prompted models using disjoint context subsets and shared prompt templates. The final prediction is then obtained by aggregating the probability vectors produced by each prompted model~\cite{duan2024privacy}.


In the tabular domain, DP has been explored primarily in the context of private synthetic data generation~\cite{carey2024dp}. This approach typically injects noise into individual records or aggregate statistics. However, such a method was designed for tabular LLMs rather than tabular FMs operating under the ICL paradigm, in which both differ mainly in data representations and inference mechanism. 

In addition to its negative impact on predictive performance, DP can also compromise the veracity of the data. We are therefore interested in methods that preserve the truthfulness and fidelity of the original data.

\section{Related Work}
The growing interest in tabular FMs has prompted researchers to investigate their potential privacy vulnerabilities. However, existing studies have primarily focused on the generative capabilities of these models. In particular, Ward et al.~\cite{ward2025tables} and Byun et al.~\cite{byun2025risk} leverage TabPFN to generate synthetic tabular data through sequential in-context prediction of features conditioned on previously generated ones. Both works demonstrate that this generator is vulnerable to MIAs. Despite the importance of these studies, our focus lies on classification tasks rather than data generation.

Regarding the exploitation of attention mechanisms for privacy attacks, this direction has been explored so far in the context of LLMs. Zaree et al.~\cite{zaree2026attenmia} propose an attention-based MIA that leverages layer- and head-level correlation statistics together with barycentric drift metrics to quantify the consistency and variation of attention patterns across transformer layers and heads. These features are combined with perturbation-based divergence metrics and used to train a dedicated attack classifier. Their results show that attention-based signals can generalise across datasets and architectures, while also identifying the layers and heads where membership leakage is most pronounced.

Despite the relevance of this work, several important differences distinguish it from our setting. First, their approach is designed specifically for LLMs and relies on an auxiliary attack classifier, whereas our objective is to develop a simpler and more direct method tailored to tabular transformers. Second, their feature extraction process depends on perturbations of the textual context, including token dropping, token replacement, and prefix insertion. Such perturbations are not naturally applicable to tabular contexts composed of structured rows rather than text sequences. In contrast, our work focuses on exploiting attention signals directly within tabular FMs for classification tasks, without requiring context perturbations or additional attack models.

\section{Tabular FMs Under Standard MIAs}
The potential leakage in tabular FMs originates from the model's dependence on the inference-time $D_{ctx}$. The resulting ICL predictor is defined as:
\[ f_{\theta, D_{\text{ctx}}}(x_q) = P_\theta(y_q \mid x_q, D_{\text{ctx}}).\]

Given black-box query access to $f_{\theta, D_\mathrm{ctx}}$ and a candidate pool $\mathcal{P}$, the adversary aims to infer whether $x_i \in D_\mathrm{ctx}$, from the behaviour of the ICL predictor on $x_i$, computing:

\[s_i = \Lambda\bigl(x_i, f_{\theta, D_{\text{ctx}}}, \mathcal{K}\bigr).\]

The adversary has no access to $D_\mathrm{ctx}$ or the model parameters. They possess an auxiliary dataset $D_\mathrm{ref}$ drawn from the same distribution as $D_\mathrm{ctx}$, used to simulate the target model's behaviour. 
The final membership prediction is obtained by thresholding the score $s_i$:
\[
\hat{m}_i
=
\mathbbm{1}[s_i > \tau].
\]
where $\tau$ is a decision threshold and $\hat{m}_i \in \{0,1\}$ indicates the predicted membership status of $x_i$ with respect to $D{\text{ctx}}$.

The adversary can be then expressed as:
\[
\mathcal{A}(x_i, f_{\theta})
=
\mathbbm{1}
\Big[
\Lambda(x_i, f_{\theta,D_{\text{ctx}}}, \mathcal{K}) > \tau
\Big] \in \{0,1\}.
\]

A tabular FM \(f_\theta\) is vulnerable to MIAs if an adversary can distinguish members from non-members with performance significantly better than random guessing. One measure of this distinguishability is the adversarial advantage:
\[
\mathrm{Adv}(\mathcal{A})
=
\left|
\Pr[\hat{m}_i = 1 \mid m_i = 1]
-
\Pr[\hat{m}_i = 1 \mid m_i = 0]
\right|,
\]
which corresponds to the gap between the true positive rate (TPR) and false positive rate (FPR) at a given decision threshold. 
Intuitively, if a candidate record \(x_i\) appears in \(D_{\text{ctx}}\), the model may assign systematically different predictive statistics to \(x_i\) relative to non-members. Such differences may emerge through confidence scores, predictive uncertainty, or attention-based conditioning effects induced by similar context rows.

In this work, we investigate whether the ICL mechanism of tabular FMs induces a detectable separation between members and non-members of \(D_{\text{ctx}}\). 

\subsection{Experimental Setup}~\label{sec:setup}
\textbf{Datasets.} We evaluate our methods on six tabular classification datasets spanning a diverse range of dataset sizes, feature dimensionalities, and class cardinalities (Table~\ref{tab:datasets}, Appendix~\ref{app:data}). They cover several data domains as well. 

\textbf{Models.} We evaluate eight classifiers across two families. We use four classical models: Multilayer Perceptron (MLP), Random Forest (RF), LightGBM, and TabNet~\cite{arik2021tabnet}. The four tabular FMs are: TabPFN~\cite{hollmann_tabpfn_2023}, Real-TabPFN, TabICL~\cite{qu_tabicl_2025}, and TabDPT~\cite{ma2026tabdpt}. Classical models are tuned with Optuna (30 trials, 3-fold CV). In contrast, FMs use fixed pre-trained weights and perform inference/adaptation through their built-in mechanism. 
More details on model architecture in present in Appendix~\ref{app:models_arch}.

\textbf{Attacks}. We evaluate three state-of-the-art MIAs:
\begin{itemize}

\item LiRA~\cite{carlini2022membership}: is a likelihood-ratio MIA that compares the target model's confidence on a sample $(x,y)$ against the confidence expected for non-member samples. In its online form, LiRA estimates both IN and OUT confidence distributions using reference models trained with and without the target sample. In the offline variant, reference models are trained independently before observing the target query and are not trained on the target sample. The attack then estimates only the OUT distribution, typically using a Gaussian approximation of confidence scores, and assigns higher membership scores to samples for which the target model's confidence is unusually large under this OUT distribution.

\item RMIA~\cite{zarifzadeh2023low}: is a relative likelihood-ratio MIA designed to reduce the cost of shadow-model attacks. Rather than scoring a sample only by its absolute confidence under the target model, RMIA compares the target sample against auxiliary population samples and uses reference models to calibrate this comparison. In the offline setting, the reference models are trained independently of the target query and can be used as OUT models. This makes the attack less dependent on training query-specific IN reference models while still retaining a likelihood-ratio interpretation.

\item Attack-P (population attack)~\cite{ye2022enhanced}: is a population-based MIA that does not train reference models. It uses auxiliary samples drawn from the same population as the target training data to estimate how the target model behaves on non-members. A target sample is predicted as a member when the target model assigns it an unusually high confidence, or low loss, relative to this population baseline.
\end{itemize}
We use a single reference model for RMIA and LiRA, since it has been shown that one model is sufficient to estimate the membership signal reliably, while avoiding the additional computational cost of training multiple references~\cite{pera2026sok,zarifzadeh2023low}. Also, we report the results concerning their offline versions, focusing on a black-box scenario.

\textbf{Data split.} For each dataset, we use a stratified 75/25 split. The 75\% portion is used as the candidate pool from which target training sets and audit samples are drawn, while the remaining 25\% is reserved as an auxiliary population set for attacks that require population data.


\textbf{Reproducibility.} All experiments are repeated across five independent random seeds. Each seed induces a new random permutation of the candidate pool, resulting in different training contexts and audit splits. Reported results correspond to the mean and standard deviation across seeds.

\textbf{Metrics.} We report the attack effectiveness using the area under the ROC curve (AUC), which provides a threshold-independent measure of separability between members and non-members (aggregate discrimination). Additionally, we report
TPR at low FPR corresponding to practical attack scenarios in which the adversary acts only on highly confident membership predictions.

\subsection{Experimental Results}


Table~\ref{tab:rmia-lira-population-locations-dropout} reports the AUC results for the MIAs evaluated in classical machine learning models and tabular FMs. Overall, the results demonstrate that both model families are vulnerable to MIAs. 
A first important observation is that tabular FMs exhibit privacy leakage levels comparable to those of classical models. In multiple settings, FMs such as TabICL and TabDPT achieve attack AUC values that are similar to, or even higher than, those observed for RF and LightGBM. For instance, on the Locations dataset, TabDPT reaches an RMIA AUC of 0.902, while for TabICL the attack success is 0.899. In Dropout Success, TabDPT achieves an RMIA AUC of 0.975, which is among the highest values observed across all evaluated models. These results show that in-context prediction with fixed pre-trained tabular transformers is still vulnerable to MIAs.

Models exhibiting larger generalisation gaps between training and test accuracy (Table~\ref{tab:train-test-accuracy-roc-datasets}, Appendix~\ref{app:targets_performance}) tend to be more vulnerable to membership inference. Although several FMs also exhibit high training accuracy, conventional overfitting alone cannot explain their privacy leakage. In particular, TabICL and TabDPT display some of the highest attack AUCs despite achieving competitive generalisation performance. For instance, TabICL obtains the highest test accuracy on Locations (0.871) while simultaneously exhibiting strong membership leakage across attacks. Similarly, TabPFN and Real-TabPFN achieve relatively modest generalisation gaps compared to classical baselines, yet still remain substantially vulnerable to MIAs. 

This distinction is particularly evident for TabDPT. Despite its strong predictive performance and relatively stable test accuracy, TabDPT remains highly vulnerable to RMIA. Even Attack-P, which lacks the model-specific calibration provided by reference models, detects a strong membership signal. This suggests that the leakage is not explained by predictive accuracy alone, but by how the model uses in-context examples and by information retained in the context representation, which can preserve sample-specific effects from the in-context training examples.

Figure~\ref{fig:mia_performance} (Appendix~\ref{app:attack_performance}) further compares the attacks in six datasets, demonstrating that RMIA is consistently the strongest attack across the two models family.

Overall, the comparison between predictive performance and attack success suggests that traditional generalisation metrics alone are insufficient to characterise privacy risk in tabular FMs. While overfitting contributes to leakage in both classical and FMs, the results indicate that tabular ICL introduces additional privacy risks that are specific to context-based inference mechanisms.

\begin{tcolorbox}[
    colback=gray!10,
    colframe=black,
    boxrule=0.5pt,
    left=1mm,right=1mm,top=1mm,bottom=1mm
]
\textbf{Key Insight ::} Tabular FMs are consistently more vulnerable to standard MIAs than classical models.
\end{tcolorbox}

\begin{table}[ht!]
\centering
\scriptsize
\include{2datasets-rmia-lira-attackp}
\caption{Attack AUC results of RMIA, LiRA and Attack-P across classical machine learning models and tabular foundation models on Locations and Dropout Success datasets.}
\label{tab:rmia-lira-population-locations-dropout}
\end{table}

\section{Attention-based MIA}
The previous results motivate us to further investigate the role of the attention mechanism in tabular FMs and its potential implications for MIAs. In transformer architectures, the attention mechanism maps a query and a set of key-value pairs to an output representation, where the query, keys, values, and output are all vector representations. The output is computed as a weighted combination of the values, with each weight determined by a compatibility function between the query and the corresponding key~\cite{vaswani2017attention}. Consequently, the attention mechanism enables the model to selectively focus on context rows that are considered more relevant to the queried sample during inference. In practice, the attention function is computed simultaneously over a set of queries, which are packed into a query matrix (Q). Similarly, the corresponding keys and values are packed into matrices (K) and (V), respectively. For a specific layer \(l\) and head \(h\), the attention weights are computed as~\cite{vaswani2017attention}:

$$
A^{(l,h)} = \mathrm{softmax}\!\left(\frac{Q^{(l,h)} {K^{(l,h)}}^\top}{\sqrt{d_k}}\right) \in \mathbb{R}^{|Q| \times |K|}
$$

The values \(V^{(l,h)}\) are then used to produce the output:
\[
O^{(l,h)}
=
A^{(l,h)} V^{(l,h)}.
\]



\subsection{Threat Model and Attack Overview}

We consider an adversary who aims to infer whether a candidate example $x_j \in \mathcal{P}$ was used as part of the context dataset $D_\mathrm{ctx}$ of a fitted target model $f_\theta$, given a candidate pool $\mathcal{P} = \{x_i\}_{i=1}^m$ and query access to $f_\theta$.

We assume a \emph{grey-box} setting: the adversary knows the architecture of $f_\theta$ but does not modify the model parameters. Notably, the adversary can patch attention modules during inference to extract attention matrices from a selected set of layer-head pairs $\mathcal{M} = \{(l,h)\}$.

For each candidate $x_j$, the adversary performs a single forward pass $f_\theta(x_j \mid D_\mathrm{ctx})$ and extracts $\{A^{(l,h)}\}_{(l,h)\in\mathcal{M}}$. Since $D_\mathrm{ctx}$ may contain auxiliary rows (e.g.\ learned \textit{thinking rows} in TabPFN), attention is restricted to the $n$ true context positions $\mathcal{I}_{\mathrm{ctx}}$ and renormalised.
Let $W^{(l,h)}$ denote the resulting submatrix of $A^{(l,h)}$ containing
attention from $x_j$ to those $n$ real context examples.
The attack then computes
\[
a_j^{(l,h)}
=
\max_{i=1,\ldots,n}
W_{j,i}^{(l,h)},
\]
which measures the strongest attention assigned by candidate $x_j$ to any
context example in $D_{\mathrm{ctx}}$.

The final membership risk score is obtained by averaging over the selected
layers and heads:
\[
s_j
=
\frac{1}{|\mathcal{M}|}
\sum_{(l,h)\in\mathcal{M}}
a_j^{(l,h)}.
\]

A candidate with a higher score is considered more likely to be a member. Evaluation is performed across all thresholds via the ROC curve. 

Algorithm~\ref{alg:amia_simple} summarises the computation of $s_j$ for all candidate examples in the attack pool.

\begin{algorithm}[ht!]
\caption{Attention-based Membership Inference Attack (AMIA)}
\label{alg:amia_simple}
\begin{algorithmic}[1]
\Require Fitted target model $f_\theta$ with context dataset $D_{\mathrm{ctx}}$
         of size $n$, attack pool $\mathcal{P}=\{x_j\}_{j=1}^{m}$, batch size $B$
\Ensure  Membership risk scores $s_1,\ldots,s_m$

\ForAll{batches $\mathcal{B}\subseteq \mathcal{P}$}
    \State Run $f_\theta$ on $\mathcal{B}$; collect
           $\{A^{(\ell,h)}\}_{(\ell,h)\in\mathcal{M}}$
    \ForAll{$(\ell,h)\in\mathcal{M}$}
        \State $W^{(\ell,h)} \gets A^{(\ell,h)}[-|\mathcal{B}|:,\;
               \mathcal{I}_{\mathrm{ctx}}]$
               \Comment{restrict to query rows and real context keys}
        \State Renormalise each row of $W^{(\ell,h)}$ 
    \EndFor
    \ForAll{$x_j\in\mathcal{B}$}
        \State $s_j \gets \dfrac{1}{|\mathcal{M}|}
               \displaystyle\sum_{(\ell,h)\in\mathcal{M}}
               \max_{i=1,\ldots,n} W^{(\ell,h)}_{j,i}$
    \EndFor
\EndFor
\State \Return $s_1,\ldots,s_m$
\end{algorithmic}
\end{algorithm}

Standard attacks operate in a black-box setting, observing only output probabilities and requiring an auxiliary dataset $D_\mathrm{ref}$ from the same distribution as $D_\mathrm{ctx}$ to calibrate membership scores. In contrast, AMIA operates in a grey-box setting, leveraging access to internal attention matrices during the forward pass while requiring no auxiliary data beyond the candidate pool $\mathcal{P}$.

\subsection{Experimental Results}
\begin{figure}[ht!]
\centerline{\includegraphics[width=\columnwidth]{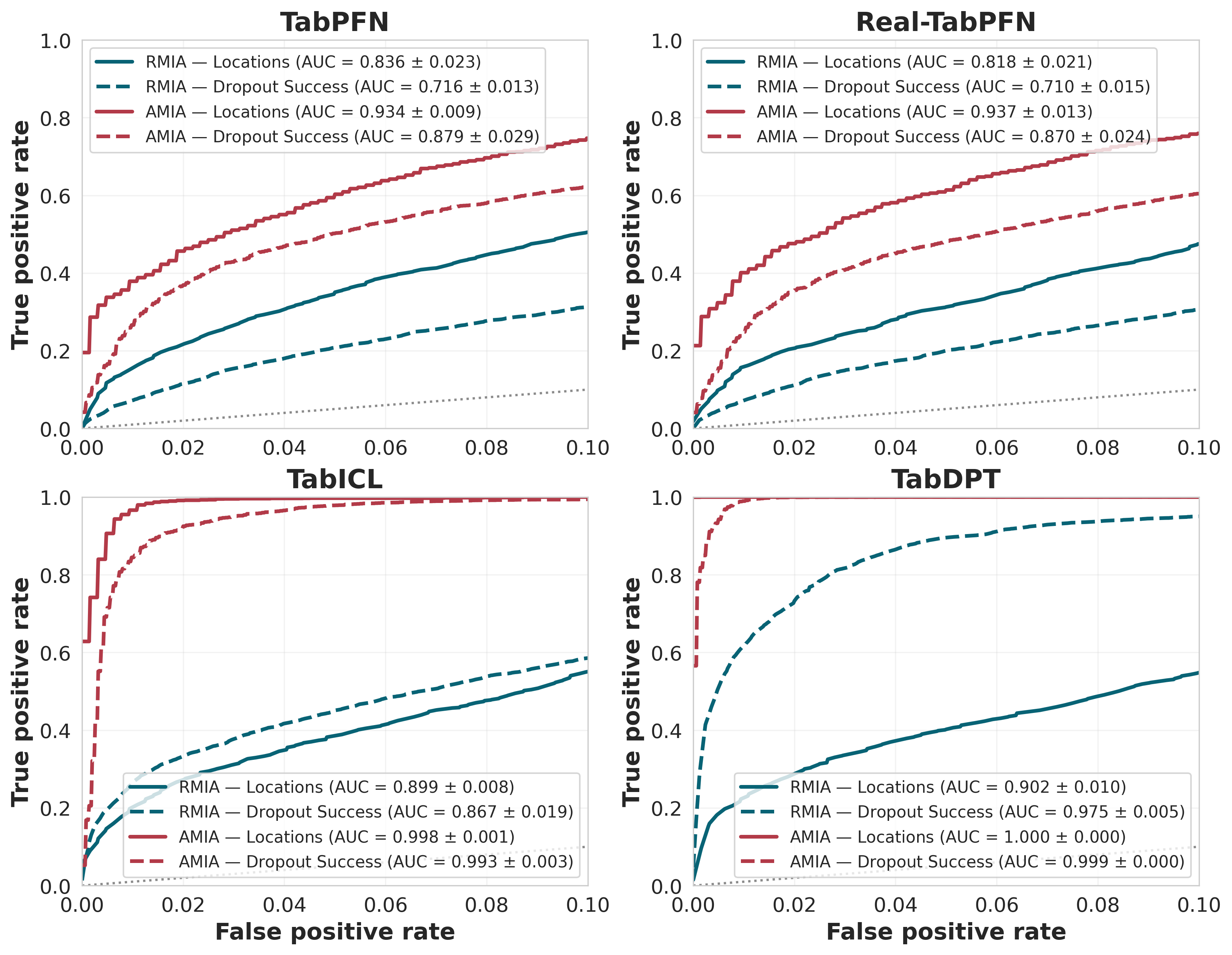}}
\caption{ROC curves (FPR $\leq$ 0.10) for AMIA and RMIA on four tabular foundation models across two datasets. Solid lines denote Locations and dashed lines Dropout Success. 
}
\label{fig:results_amia}
\end{figure}

The following results focus on the evaluation of the four tabular FMs, since AMIA is not applicable to classical machine learning models. Based on previous best results, we also focus on RMIA comparisons only. The setup of all experiments remains the same as described in Section~\ref{sec:setup}. Figure~\ref{fig:results_amia} depicts the TPR at low FPR for both attacks.

We observe that AMIA consistently outperforms RMIA across all models and both datasets. In particular, the attention-based signal yields substantially stronger MIA performance than the output-based signal used by RMIA, indicating that attention weights expose membership information that is not captured by prediction probabilities alone.

For TabPFN and Real-TabPFN, AMIA improves the AUC by approximately 10 percentage points over RMIA on Locations. For TabICL and TabDPT, the difference is considerably more pronounced. RMIA achieves an AUC of 0.899 and 0.902 for these models respectively, whereas AMIA achieves near-perfect discrimination with an AUC of 0.998 and 1.000. This suggests that such architectures learn sharp and discriminative attention patterns toward memorised context examples.

Among all evaluated models, TabDPT appears to be the most vulnerable. AMIA achieves an AUC of approximately 100\% on both datasets, implying that membership can be inferred almost perfectly from attention information alone. Correspondingly, the ROC curve is almost vertical at very low FPR, showing that the adversary can identify the vast majority of members before any false alarms occur.

Finally, Real-TabPFN exhibits behaviour that is almost identical to that of TabPFN. Including real-world data has little effect on attack performance, suggesting that the observed vulnerability is primarily architectural rather than a consequence of the pre-training data. 

Additionally, when considering $FPR=0$, we observe that RMIA collapses to near zero across all models demonstrating that cannot reliably identify any member without also flagging non-members. On the other hand, AMIA retains meaningful signal. For example, on Locations, TabPFN AMIA identifies 19\% of members with zero false alarms and TabICL AMIA notably identifies 63\%.

A key difference between both TabPFN versions and TabICL and TabDPT is the presence of thinking rows. 
Removing thinking rows from TabPFN and Real-TabPFN it prevents non-context data from being counted as training context attention. However, thinking rows still participate in the forward pass and can influence hidden representations through propagation across layers. As a result, they may dilute the membership signal. 

\begin{tcolorbox}[
    colback=gray!10,
    colframe=black,
    boxrule=0.5pt,
    left=1mm,right=1mm,top=1mm,bottom=1mm
]
\textbf{Key Insight ::} The attention concentration signal is a stronger and more consistent membership indicator than output-based signals with AMIA approaching near-perfect discrimination while requiring no population data or reference model training.
\end{tcolorbox}

\section{Inference-Time Hardening Against MIAs}
The AMIA results show that membership leakage in tabular FMs is not limited to output probabilities. In many settings, the most notable signal originates from the model's internal attention patterns where members frequently generate highly focused attention on the training context. This suggests that attention can be an additional source of privacy vulnerability. 
Methods such as HAMP~\cite{chen2023overconfidence} or restricting the prediction vector to the top-$c$ classes~\cite{truex2019demystifying}, which are effective against classical confidence-based attacks, do not modify the attention patterns exploited by AMIA. Similarly, differentially private synthetic data generation~\cite{carey2024dp} could replace the private context with a more protected version, but this approach requires training an additional generative model and completely discards the original context, thus undermining the ICL mechanism that makes FMs effective.

Therefore, we propose a defence specifically designed for the tabular ICL setting. Rather than modifying the model through retraining, our approach operates directly at inference time by modifying the row-attention mechanism to reduce the discriminability of context-key representations.

\subsection{High-risk Attention Control}
When designing a defence mechanism, a key objective is to preserve predictive performance. This can be achieved more effectively by observing that not all queries exhibit the same level of privacy risk. In particular, member queries tend to produce high attention scores ($s_j$), as the model has successfully memorised their corresponding context keys. In contrast, non-member queries typically induce more diffuse attention distributions, reflecting weaker and less specific associations within the context.

Given the AMIA scores, we therefore flag only the subset of queries deemed high-risk:
\[
  \mathcal{H}_\tau = \{x_j \in \mathcal{B} : s_j \geq \tau\},
\]
Queries in $\mathcal{B} \setminus \mathcal{H}_\tau$ are served with the original, unmodified context keys, preserving accuracy on the majority of inputs.

The threshold $\tau$ controls the privacy-utility trade-off and can be set in two ways. If the practitioner specifies $\tau$ directly (e.g. derived from domain knowledge or a desired fallback rate), it is used as-is. Otherwise, $\tau$ is calibrated from a reference set using TPR-based calibration. In the latter case, since the defender has full knowledge of $D_{\mathrm{ctx}}$, the threshold is calibrated by computing AMIA scores on the context examples themselves. Specifically, $\tau$ is set to the minimum score among context members:
\[
  \tau = \min_{x_j \in D_{\mathrm{ctx}}} s_j,
\]
ensuring that any query whose attention concentration is at least as sharp as the least-exposed context example is flagged as high-risk. This allows to apply the anonymisation exclusively to those queries.

Our defence uses $k$-anonymity~\cite{Sweeney2002kAnonymityAM} and microaggregation~\cite{DomingoFerrer2002PracticalDM}, a privacy-preserving data publishing approach. The goal of $k$-anonymity is to ensure that each released record is indistinguishable from at least \(k-1\) other records with respect to a chosen set of quasi-identifying attributes. In other words, a protected record should belong to a group, rather than appearing as a unique individual. Microaggregation is a well-known method for achieving $k$-anonymity, where similar records are clustered into small groups, and replaced by a representative aggregate, such as the group centroid. This reduces record-level uniqueness while preserving data structure.

Therefore, instead of protecting released database records, our objective is to prevent individual context examples from becoming uniquely identifiable attention keys. To achieve this, we employ label-aware microaggregation, where each record is grouped with at least \(k-1\) nearby examples from the same class, and the model is given an aggregated representative instead of the original individual record. This directly targets AMIA's leakage mechanism by reducing sharp attention to isolated context examples, while the label constraint helps preserve predictive utility.

Algorithm~\ref{alg:highrisk_kanon} specifies all the steps for this targeted, label-wise microaggregation approach.

\begin{algorithm}[ht!]
\caption{Targeted Label $k$-Anonymity}
\label{alg:highrisk_kanon}
\begin{algorithmic}[1]
\Require Fitted target model $f_\theta$ with context dataset $D_{\mathrm{ctx}}$
         of size $n$, context labels $\{y_{\mathrm{ctx},i}\}_{i=1}^{n}$,
         AMIA scores $\{s_j\}$ for query batch $\mathcal{B}=\{x_j\}$,
         anonymity parameter $k$,
         smoothing parameter $\alpha \in [0,1)$,
         threshold $\tau$
\Ensure  Predictions $\hat{p}_1,\ldots,\hat{p}_{|\mathcal{B}|}$

\ForAll{class $c \in \mathcal{Y}$}
    \State $\mathcal{G}_c \gets \{i : y_{\mathrm{ctx},i} = c\}$;\quad
           \textbf{skip} if $|\mathcal{G}_c| < k$
    \State Partition $\mathcal{G}_c$ into non-overlapping groups of size $k$
           by key similarity
    \ForAll{groups $g$}
        \State $\mu_g \gets \dfrac{1}{|g|}\displaystyle\sum_{i \in g} K_i$
        \ForAll{$i \in g$}
            \State $\tilde{K}_i \gets \alpha\, K_i + (1-\alpha)\,\mu_g$
        \EndFor
    \EndFor
\EndFor

\ForAll{$x_j \in \mathcal{B}$}
    \If{$s_j \geq \tau$}
        \State $\hat{p}_j \gets f_\theta(x_j \mid D_{\mathrm{ctx}},\, \{\tilde{K}_i\})$ \Comment{forward pass with anonymised keys $\tilde{K}$}
    \Else
        \State $\hat{p}_j \gets f_\theta(x_j \mid D_{\mathrm{ctx}})$ \Comment{normal forward pass}
    \EndIf
\EndFor

\State \Return $\hat{p}_1,\ldots,\hat{p}_{|\mathcal{B}|}$
\end{algorithmic}
\end{algorithm}

Before any query is processed, the context keys are anonymised class-conditionally. For each class $c \in \mathcal{Y}$, the set of context indices with label $c$ is collected as
$\mathcal{G}_c = \{i : y_{\mathrm{ctx},i} = c\}$. Classes with fewer than $k$ examples are skipped, as a group of size $k$ cannot be formed.

Within $\mathcal{G}_c$, context examples are sorted by key similarity and partitioned into non-overlapping groups $g$ of size $k$. Grouping is performed within the same class to avoid mixing examples of different labels, which would corrupt the attention-weighted predictions. For each group $g$, the centroid of the key vectors is computed as
\[
  \mu_g = \frac{1}{k}\sum_{i \in g} K_i,
\]
and each key is replaced by a convex interpolation between its original value and the group centroid:

\[
  \tilde{K}_i = \alpha\, K_i + (1-\alpha)\,\mu_g, \quad \alpha \in [0,1).
\]
While the original microaggregation replaces keys entirely with a centroid, soft label microaggregation allows for small perturbations in predictions.
For $\alpha=0$, all keys in a group share the same centroid representation, yielding $k$-anonymous groups in key space. For $\alpha>0$, the mechanism interpolates between the original and anonymised keys, trading privacy for utility. The parameter $\alpha$ thus directly controls the anonymisation strength: a lower value makes it harder for an adversary to distinguish which specific context example a query is attending to, at the cost of reduced attention precision and potential accuracy loss.


At inference time, each query $x_j \in \mathcal{B}$ is routed based on its AMIA score. If $s_j \geq \tau$, the forward pass uses the anonymised keys $\tilde{K}_i$, replacing the original context-key representations in all context-attention layers $(\ell,h)\in\mathcal{M}$.
If $s_j < \tau$, the original forward pass $f_\theta(x_j \mid D_{\mathrm{ctx}})$ is returned unchanged. This selective strategy ensures accuracy for low-risk queries while preventing AMIA from identifying the most relevant ones.






\subsection{Experimental Results}
In this section, we investigate the effectiveness of the target label $k$-anonymity defence. We first study the influence of the threshold determined by the minimum known members score, followed by an analysis of the attention transformation applied to the selected queries. All experimental settings remain unchanged from the previous section.

\begin{figure*}[ht!]
\centerline{\includegraphics[width=0.9\textwidth]{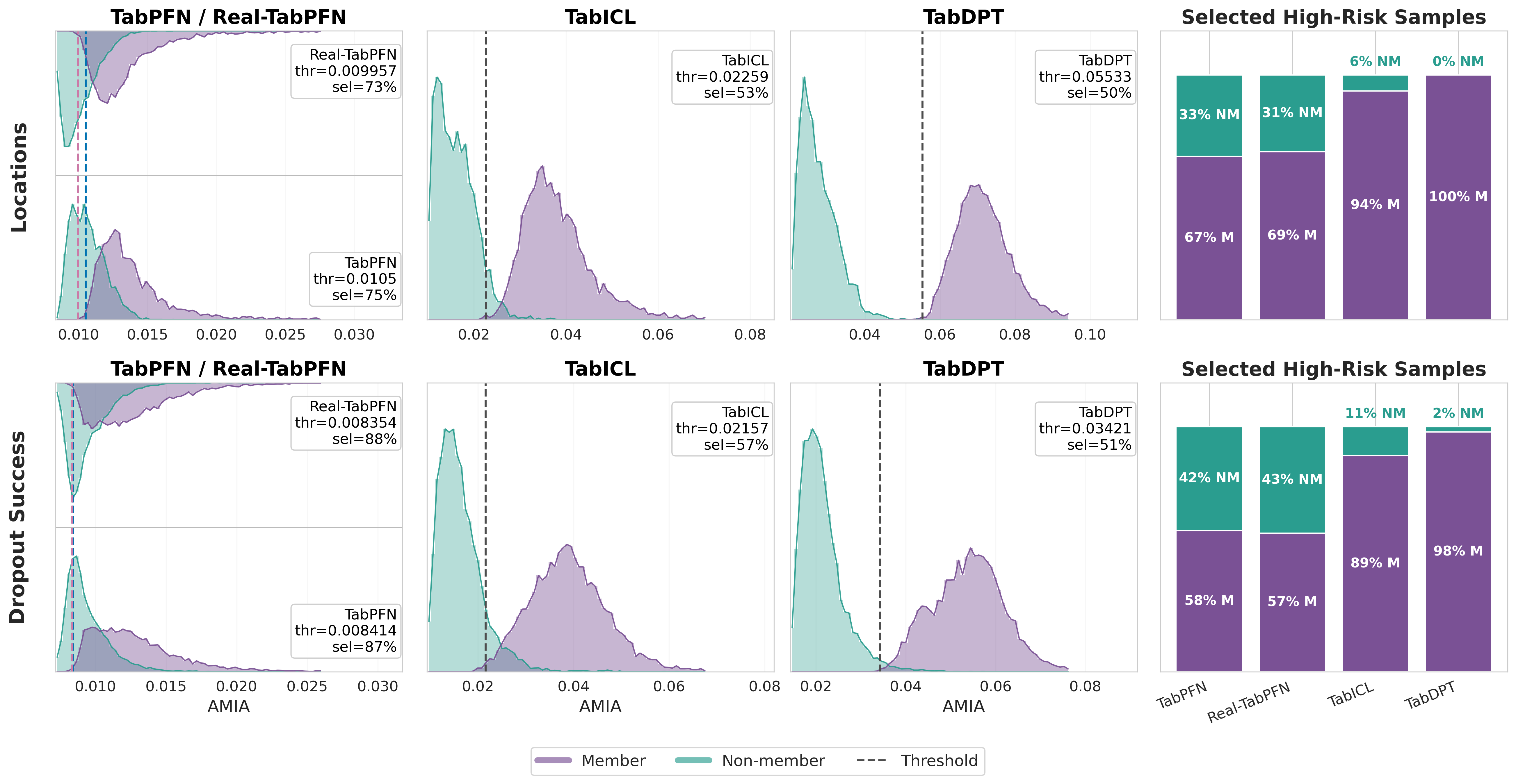}}
\caption{AMIA-based high-risk selection. The density plots show the full distribution of AMIA row-attention scores for all member and non-member samples, with the dashed line indicating the selection threshold. The stacked bar plots summarise the samples above this threshold, showing the member/non-member composition of the selected high-risk subset.} 
\label{fig:highrisk_selection}
\end{figure*}

\textbf{High-risk selection. }
The first results concern Phase 2, where we evaluate the threshold selection for high-risk query identification. Figure~\ref{fig:highrisk_selection} illustrates the discriminative capacity of AMIA scores in identifying true members across tabular FMs. The dashed vertical line denotes the learned threshold $\tau$, above which queries are flagged as high-risk and assigned to the target label $k$-anonymity defence.

The results suggests that AMIA is much more membership discriminative for TabDPT and TabICL in which member rows receive substantially higher maximum attention scores than non-members, creating a clear distributional separation. For TabDPT, the member and non-member score distributions are nearly disjoint in which the threshold falls cleanly between them, flagging exclusively members with a selection rate of 50\%. TabICL shows similarly good separation with a selection rate of 53\% and 57\%, meaning a small but non-negligible fraction of non-members are caught by the threshold. In both cases the defence will behave as intended by intervening on a targeted high-risk subset while leaving the remaining queries unaffected.

On the other hand, both TabPFN and Real-TabPFN exhibit poor separation. The member and non-member AMIA score distributions overlap heavily and concentrating near zero. In these cases, the threshold provides little discriminative power. Despite an AMIA AUC of 93\%, the selection rate reaches 75\% on Locations and 87\% on Dropout Success, with member shares of only 67\% and 69\%. The high AMIA AUC observed in Figure~\ref{fig:results_amia} reflects membership leakage, but the diffuse attention concentration signal in TabPFN is spread across many context rows rather than sharply concentrated on a few.

As evidence, Figure~\ref{fig:attack_layers} (Appendix~\ref{app:layers_performance}) reports the per-layer AMIA AUC. TabPFN has zero layers that achieve near-perfect separation. The per-layer AUC reaches at most 0.98 on Locations and 0.90 on Dropout Success, with the signal oscillating across all 24 layers and never concentrating cleanly in any subset. Therefore, while the aggregate AMIA score produces a high ROC AUC, the underlying score distributions remain overlapping. In particular, the non-member right tail and the member left tail occupy similar score ranges. As a result, any threshold set to capture a meaningful proportion of members will inevitably fall within the non-member distribution, explaining the high non-member selection rate observed.

For TabDPT, 7 out of 16 layers on Locations achieve a per-layer AUC above 0.99. When the final AMIA score aggregates these layer signals, the resulting distribution keeps this sharp separation, which is why the same high recall threshold selects almost exclusively members.
Such results may be attributed to architectural design. TabPFN runs many fragmented attention calls per layer, each covering only a subset of features, which may cause diffuse signals. In contrast, TabDPT uses a single call per layer over a complete row representation, with key vectors encoding all features simultaneously. Thus, its attention signal can concentrate more directly on the member rows.

\textbf{Target label $k$-anonymity. } Next, we evaluate the effectiveness of the proposed defence in mitigating membership leakage on the previously identified high-risk queries. We consider following parameters inn our defence: $\alpha \in \{0,0.3,0.5\}$ and $k \in \{3,5,10\}$.

\begin{figure}[ht!]
\centerline{\includegraphics[width=\columnwidth]{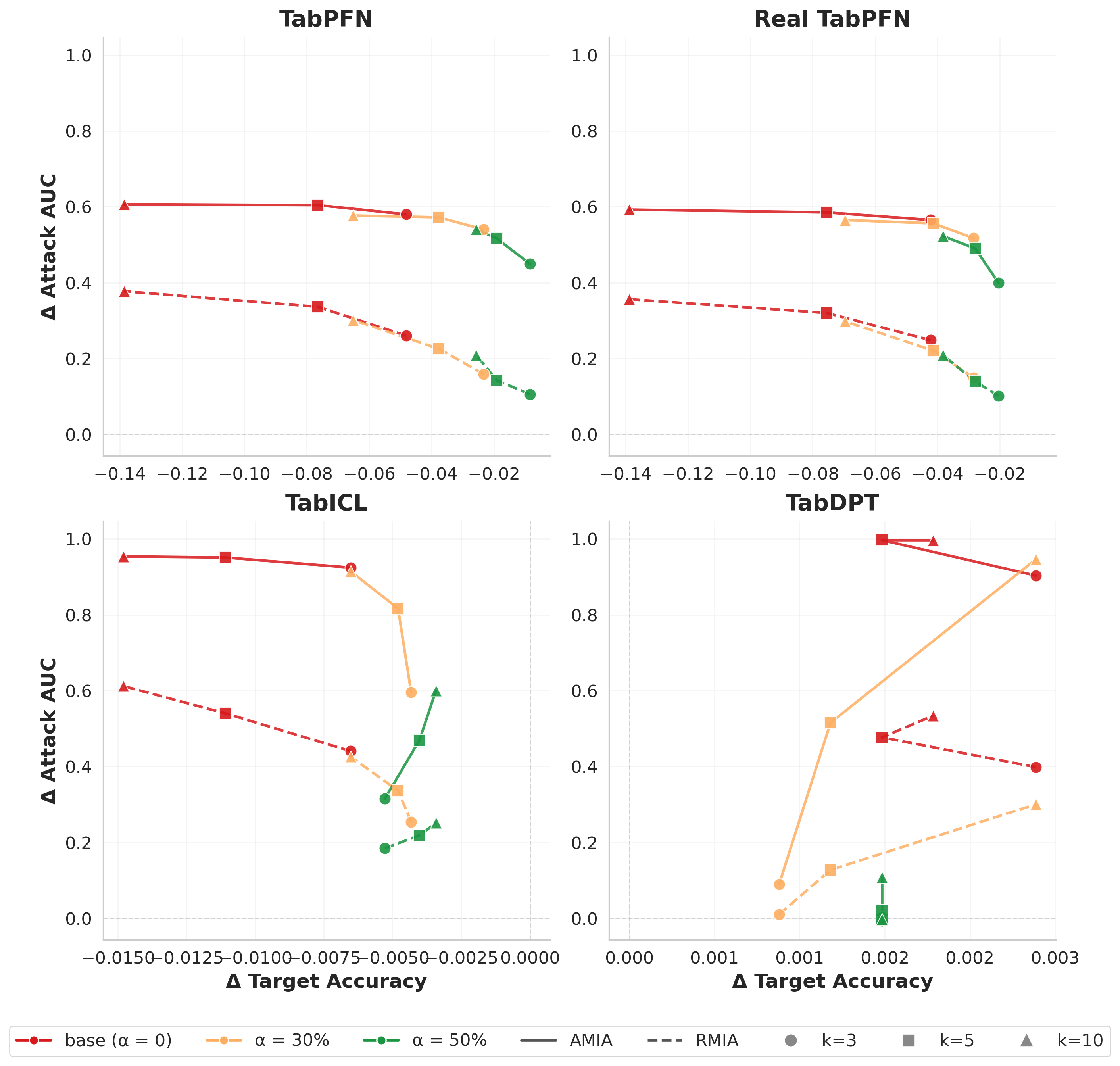}}
\caption{Privacy-utility trade-off of targeted label $k$-anonymisation defence averaged over Locations and Dropout Success datasets.} 
\label{fig:high-risk-kanon}
\end{figure}

Figure~\ref{fig:high-risk-kanon} shows the change in the attack AUC versus the change in accuracy for each defence configuration. The ideal point is at upper-right, which corresponds to privacy gain and minimum accuracy degradation. Each curve connects configurations of the same $\alpha$ across $k$ values.

We observe a large accuracy cost and considerable privacy gain for TabPFN and Real-TabPFN. The defence reduces AMIA AUC from 0.40 and up to 0.61 and RMIA by 0.10 to 0.38, but at higher accuracy cost. Base target label $k$-anonymity ($\alpha$=0\%) decreases accuracy by up to 14\%. This is a direct consequence of high-risk selection, as many non-members are being transformed too. Softening to $\alpha=50\%$ reduces accuracy cost substantially but also weakens the privacy gain specially for RMIA. This clearly highlights a trade-off. However, the highest softened $\alpha$ allows for a reduction of AMIA around the random guess with practically no costs for predictive utility. On the other hand, this is not possible for RMIA, which requires stronger parameters to increase privacy gain. This means that the defence targets the attention signal more directly than the output confidences.

In the case of TabICL, base target label $k$-anonymity reduces AMIA AUC by up to 95\% with an accuracy drop of only 1.5\%. Even with $\alpha=50\%$ can yield 0.19 and 0.60 AMIA reduction with negligible accuracy impact. The small accuracy cost reflects the precise selection of high-risk queries; the defence is applied mostly to true members, so non-member predictions are largely unaffected. This is the regime in which adaptive defence works as intended.

For TabDPT, a nearly perfect reduction in AMIA can be achieved at a negligible accuracy cost for $\alpha=0$. However, protection collapses sharply as $\alpha$ increases; at $\alpha=50\%$, the AMIA reduction is nearly zero. This is because TabDPT's attention is sharply concentrated on the member's own key, so even moving it 50\% towards the centroid still leaves it as the dominant key. This means that the member's key is far enough from all other keys that partial smoothing is insufficient to break the concentration. Full key replacement ($\alpha=0$) is required to destroy the unique signature.

A non-monotonic pattern is also observed for the $k$ parameter in TabICL and TabDPT. While larger $k$ is expected to increase the accuracy cost, the ordering reverses at $\alpha=50\%$, where $k=10$ incurs less accuracy cost than $k=3$.
This occurs because centroid stability becomes the dominant factor at soft $\alpha$.
With $k=3$, the centroid is estimated from few rows and can have high variance, so blending with it may move the key in an irregular direction. With $k=10$, the centroid is more stable and closer to the class mean, producing a smoother and less disruptive displacement. Accuracy differences are nevertheless very small (order of $10^{-3}$), as the defence is applied almost exclusively to members and the centroid prediction typically preserves the correct class regardless of $k$.

\begin{tcolorbox}[
    colback=gray!10,
    colframe=black,
    boxrule=0.5pt,
    left=1mm,right=1mm,top=1mm,bottom=1mm
]
\textbf{Key Insight ::} The AMIA risk score serves as an effective signal for both attack and defence. By selectively applying label $k$-anonymity to high-risk queries, attention- and confidence-based attacks are substantially mitigated while preserving predictive performance.
\end{tcolorbox}

\section{Fine-tuning Leakage}
While the previous sections focused on privacy leakage arising from inference-time context interactions, tabular FMs may also introduce additional risks when adapted to predictive tasks through fine-tuning. Fine-tuning is commonly used to specialise a pre-trained model to the distribution, feature space, and label semantics of a target dataset, often yielding stronger predictive performance.

However, updating model parameters on private data may also cause the model to encode information about the fine-tuning samples. As a result, fine-tuned samples can become more distinguishable from unseen data, increasing their vulnerability to MIAs. Evaluating fine-tuning is therefore important not only from a utility perspective, but also from a privacy perspective.
For demonstration purposes, we focus on a single model.
We aim to answer the question:
\textit{Does fine-tuning TabDPT on a dataset increase its MIA vulnerability for those fine-tuned training samples?}

For this, we introduce an explicit fine-tuning stage. Let the fine-tuning data be
$D_{\mathrm{ft}}=\{(x_i,y_i)\}_{i=1}^N$. 
Starting from the pre-trained parameters $\theta_0$, we update the model using gradient-based optimisation on $D_{\mathrm{ft}}$, obtaining
$$\theta_{\text{ft}} = \arg\min_\theta \mathcal{L}(\theta; D_{\text{ft}})$$

The fine-tuned model then predicts $P_{\theta_{\text{ft}}}(y_q \mid x_q, D_{\text{ctx}})$.

For this particular experiment, we consider disjoint sets
$D_{\text{ft}} \cap D_{\text{ctx}} = \emptyset$.
This ensures that any leakage observed in $P_{\theta_{\text{ft}}}$ is attributable to weight adaptation from $D_{\text{ft}}$, not to $x_q$ appearing as a context example.

For each audited sample $(x_q,y_q)$, we use the pre-trained TabDPT $\theta_0$ as a public reference model and compare it with the fine-tuned $\theta_{\mathrm{ft}}$ under the same fixed context:
\[
P_{\theta_0}(y_q \mid x_q, D_{\mathrm{ctx}})
\quad \text{and} \quad
P_{\theta_{\mathrm{ft}}}(y_q \mid x_q, D_{\mathrm{ctx}}).
\]
Then, the goal is to infer whether $(x_q,y_q) \in D_{\mathrm{ft}}$.

\textbf{Fine-tuned data leakage. }
We measure the fine-tuning membership signal using the change in true label. Intuitively, fine-tuning causes the model to assign higher confidence to the true labels of samples on which it was trained, making this shift a natural membership indicator.

\begin{figure}[ht!]
\centerline{\includegraphics[width=\columnwidth]{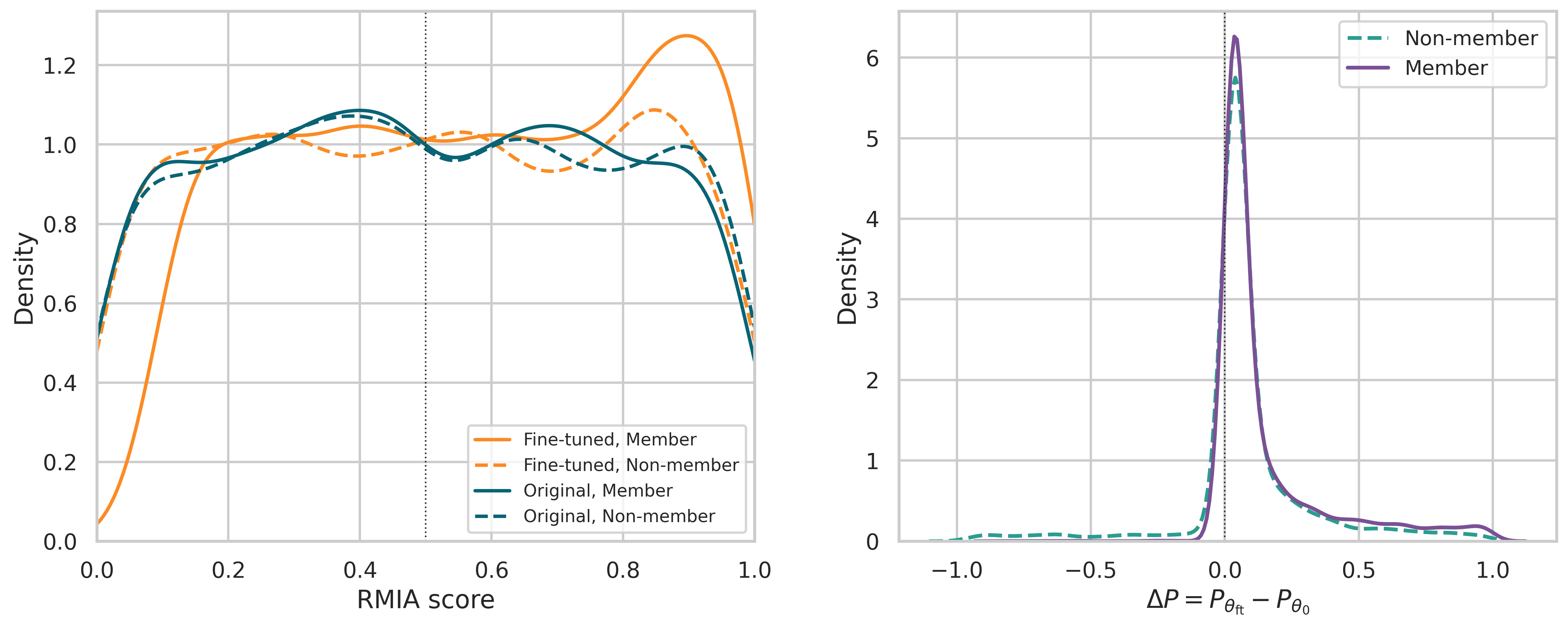}}
\caption{Fine-tuning membership signal in TabDPT using Purchases10 dataset. RMIA score distributions for fine-tuning (left) and distribution of the true label confidence shift (right).} 
\label{fig:ft_results}
\end{figure}

We report in Figure~\ref{fig:ft_results} the results for RMIA, as AMIA was designed for ICL context membership. We explore the shifts in confidences tested in Purchases10 dataset.

Before fine-tuning, RMIA is essentially at chance with AUC of 0.49 as $D_{ctx}$ does not contain any query from the $D_{ft}$. After fine-tuning, RMIA AUC rises to about 0.57 (Figure~\ref{fig:ft_train}, Appendix~\ref{app:finetuning}). This means fine-tuned training samples become more distinguishable from held-out non-members after the weight update.
In the confidence shift (right image) we observe that  members receive a larger increase in true label confidence than non-members. A rightward shift for members indicates that fine-tuning increases confidence more on samples used for weight updates. 

\textbf{Fine-tuned context leakage. }
We also evaluate the attack in context after the fine-tuned model. The adversary never sees fine-tuned samples; the idea is to distinguish between context members and non-members, isolating the privacy risk of in-context exposure from the privacy risk of weight memorisation. In this scenario, we include AMIA. Table~\ref{tab:context_mia_auc} reports the results for both attacks by comparing the leakage between the original and the fine-tuned models.

\begin{table}[ht!]
\centering
\include{finetuned}
\caption{Context MIA AUC on fine-tuned TabDPT.}
\label{tab:context_mia_auc}
\end{table}

In the original model, the membership signal detected by RMIA comes from a gap in output probability. When \(x_q\) is a context member, the model can rely directly on the matching context row to predict \(y_q\), leading to a higher true label probability. After fine-tuning, this gap is largely reduced. As a result, output confidence is no longer strongly discriminative. In particular, RMIA reaches an AUC of 0.54, close to random guessing. Even with a strong reference model, there is little remaining probability gap to exploit.

AMIA, on the other hand, reaches very high AUC for both the original and fine-tuned models. This is because AMIA does not measure output confidence. Instead, it measures where the model focuses its attention, which still concentrates on $x_q$ when it is in the context, regardless of whether the model needed to focus there in order to make the prediction. Although fine-tuning updates the model's weights it does not change the fact that context members are attended more sharply; this makes AMIA more efficient.

In summary, we confirm that \textit{i)} fine-tuning introduces a measurable, albeit modest, vulnerability on top of the in-context signal already present in TabDPT, and \textit{ii)} the confidence shift score outperforms the raw fine-tuning model's confidence as a membership discriminator, indicating that what matters is not the absolute confidence but how much it changed due to fine-tuning.

\section{Discussion}
\subsection{MIAs on tabular FMs}
Confidence-based attacks, such as RMIA, observe the model at its output boundary. Although the signal is legitimate, it can be misleading. A non-member lying far from the decision boundary in a dense, well-represented region of the feature space produces high output confidence for the same reason as a memorised training example. Therefore, output probability cannot distinguish between the two cases.

By contrast, AMIA operates on a different layer of computation. In an ICL model, the query not only makes a confident prediction but also retrieves a specific set of context rows via the attention mechanism. When the concentration score is high, the model has focused its attention on one or a few rows, effectively treating the query as almost identical to a specific context example and adopting its label. This makes memorisation structurally visible  as the model is not only confident, but also points to a specific row.

In general, AMIA adds attack power beyond RMIA. 
We also compared with the online version of RMIA (Figure~\ref{fig:off_on_performance}, Appendix~\ref{app:amia_performance}). 
RMIA-online improves over offline version, but the gain is smaller than AMIA's gain. The online boost is largest for TabPFN and Real-TabPFN, smaller for TabICL, and almost negligible for TabDPT.

Additionally, Figure~\ref{fig:runtime} (Appendix~\ref{app:computational_costs}) shows that AMIA is not only more effective, but also substantially more computationally efficient than existing attacks. In median, AMIA requires less than one minute to extract and compute attention-based membership signals. In contrast, RMIA incurs higher computational overhead, as it requires repeated scoring of the target model, reference models, and population behaviour distributions. By directly leveraging internal attention representations obtained during inference, AMIA avoids these expensive auxiliary computations while still achieving stronger attack performance.

AMIA also demonstrates to be persistent as $D_{ctx}$ changes.
Figure~\ref{fig:context_size} (Appendix~\ref{app:context_size}) shows that membership leakage is robust to the choice of context size. AMIA remains strong from small context fractions to full $D_{ctx}$. 
For RMIA, increasing the context size tends to dilute the membership signal captured by output probabilities. This indicates that RMIA is sensitive to the composition and size of $D_{ctx}$, and may underestimate membership leakage when the signal is expressed internally through attention rather than externally through prediction confidence.

Despite the generally high attack performance on all tabular FMs, their architecture influences their vulnerability. TabPFN and TabICL rely solely on synthetic data pre-training. Real-TabPFN builds on TabPFN and further trains on real datasets with the goal of aligning synthetic priors with real tabular distributions. TabDPT incorporates real-world data during pre-training. In this regard, we did not observe substantial differences between TabPFN and Real-TabPFN. Beyond pre-training, their attention mechanisms also differ. Both TabPFN variants attend directly to context rows and labels. TabICL compresses rows into learned embeddings before performing ICL, which may yield cleaner row-level representations and thus sharper attention patterns. TabDPT pre-trains a transformer encoder that learns discriminative feature representations, without explicit retrieving context. The strong attack performance on TabDPT suggests that attention-based leakage is not limited to architectures with explicit contextual retrieval, but can also emerge from highly discriminative representation learning.

\subsection{Defence}
As previously observed, the proposed target label $k$-anonymity defence is effective. Figure~\ref{fig:defences_ablation} (Appendix~\ref{app:defence_ablation}) further illustrates this by evaluating the privacy-utility trade-off of such approach against two other strategies. First, we uniformly apply label $k$-anonymity to all queries regardless of their risk score. Second, we apply attention dropout which regularises the row-attention weights at inference time by randomly zeroing out attention entries. 
While they eliminate the need for a risk scorer, they apply the defence indiscriminately to every query, including non-members. Attention dropout is the least efficient strategy for AMIA. Furthermore, we notice a higher advantage in selecting high-risk queries: simple label $k$-anonymity only reduces the attack effectiveness by a small amount.
Future work may include applying the proposed high-risk query selection strategy to each approach individually, but also their combination to obtain improved results. 

Our defence is specifically for tabular ICL. Label $k$-anonymity exploits the structured, row-level and is therefore natural for tabular ICL because each sample is a structured feature-label pair, and same-label rows can be averaged or blended in feature space while preserving the class. In text-based ICL, samples are discrete natural language sequences; there is no simple label-conditioned centroid that remains a valid, semantically faithful text example. Therefore, this defence is specific to tabular FMs and does not directly apply to text ICL. On the other hand, attention dropout can in principle be applied to text-based ICL as it operates on attention weights, not on the data format.

\subsection{Limitations and opportunities}
In our current setup, we aggregate representations by averaging across all layers. While this provides a simple and stable baseline, alternative layer-aggregation strategies should be explored in future work, as different layers may encode task-specific information. Potential directions include selecting individual layers or combining only subsets of layers most relevant to the downstream task.

Concerning our defence, it can be applied either before deployment or adaptively at inference time (real-time). In the static setting, the groups and centroids are computed only from the fixed context keys and used for all queries. In the adaptive setting, the same transformation is applied only when the high-risk guardrail identifies a sensitive query. 
If the model is defended before deployment, then the adversary cannot bypass the transformation by avoiding the trigger; they interact only with the defended model. The attack needs to know at least three parameters ($\tau$, $k$, $\alpha$) to bypass the defence. In that setting, target label $k$-anonymity is stronger. However, leakage may persist through residual output-confidence signals, group-level membership information such as minority groups, or imperfect anonymisation when the centroid still encodes sensitive local structure.

In real time, an adversary can search for query versions that preserve membership information while falling below the high-risk threshold, thereby avoiding the defence. Since the defence is applied only after detection, false negatives are the main attack surface. This limitation is inherent to hard thresholding as small changes around the decision boundary can alter whether the defence is activated. A possible mitigation is to perform detection on a local neighbourhood of the query, rather than only on the submitted query. For example, the defender could evaluate the risk score over perturbations or nearest neighbours of \(x_q\), and activate the defence if any nearby point exceeds the threshold.

A second limitation is the choice of threshold. Setting \(\tau\) as the minimum AMIA score among known members is  conservative. A low threshold may effectively protect members, but this comes at the cost of causing many non-member queries to be processed by the stronger defence.

A more robust alternative may be quantile-based member calibration. Instead of requiring all known members to be caught, we choose a threshold that catches a target fraction \(\beta\) of members: 
\(\tau_{\beta}
=
\mathrm{Quantile}_{1-\beta}
\bigl(\{s_i : x_i \in \mathcal{D}_1\}\bigr).\) 

For example, setting \(\beta = 0.9\) means that the defence is calibrated to catch approximately \(90\%\) of members, while tolerating the lowest-risk \(10\%\). 
This avoids letting a single low-score member dominate the threshold. 
The limitation is that some members are intentionally left unflagged, so the defence no longer guarantees full member coverage. Future work includes such analysis and adjustment of the threshold.

A final limitation concerns scalability. Both AMIA and target label $k$-anonymity introduce scalability challenges that constrain their practical applicability.
For AMIA, the main computational bottleneck is storing attention weights. Depending on the implementation, the captured attention tensor scales with the number of query-context pairs, and can become quadratic when full row-attention over the context/query sequence is retained.

Target label $k$-anonymity faces a different class of scalability constraints, namely when high-risk labels are rare. Achieving $k$-anonymity for under-represented classes requires aggressive generalisation which amplifies utility. In high-dimensional feature spaces, forming genuinely indistinguishable groups of size $k$ becomes increasingly difficult, as the probability that $k$ records share sufficiently similar group values decreases with dimensionality. Finally, when the inference context is dynamic, the \(k\)-anonymous partition must be recomputed or updated at inference time, introducing a computational overhead that undermines deployment feasibility. This limitation becomes especially important in deployment settings. Since the defender does not know future queries in advance, and the inference context may depend on each query, high-risk detection and \(k\)-anonymous grouping cannot always be fully precomputed and may require additional online computation.

\section{Conclusion}
In this paper, we demonstrate that tabular foundation models (FMs) despite being pre-trained on large collections of synthetic data are not free from privacy risks. First, we show that such models are sensitive to standard MIAs that are based on confidence scores. As tabular FMs operate through in-context learning, their predictions depend directly on the context examples provided at inference time, in which the transformers' attention mechanism amplifies the membership signal. Second, we propose an attack that exploits such attention dynamics called AMIA. Results show a high efficiency of AMIA compared to confidence-based attacks.
Third, we design a new defence strategy to tackle this problem based on $k$-anonymity principles by targeting high-risk queries. Our defence reduces the granularity of the context attention representations through label-aware microaggregation. Such approach reduces AMIA but also confidence-based attacks while preserving predictive utility. 
Furthermore, we demonstrate that fine-tuning can introduce additional privacy risk even when context-based attention signals remain weak.

In short, our research contributes towards the trustworthy deployment of tabular FMs, by highlighting the importance of privacy risks in-context learning systems. Our findings demonstrate that synthetic pre-training alone is insufficient to guarantee privacy, and that both inference-time contextual interactions and task-specific parameter adaptation constitute critical sources of membership leakage. More broadly, we aim for this work to pave the way for new research into the privacy risks, as such models continue to be adopted across sensitive real-world domains.






%



\bibliography{refs}
\bibliographystyle{IEEEtran}




\appendices  
\section{Datasets}~\label{app:data}
The characteristics of all datasets used to evaluate the MIAs are presented in Table~\ref{tab:datasets}. We use datasets that span a range of sample sizes, features, and class labels. The domains include healthcare, finance, education, locations and purchases. 

\begin{table}[ht!]
\centering
\scriptsize
\include{datasets}
\caption{Summary characteristics of the datasets used. }
\label{tab:datasets}
\end{table}


\section{Model architectures}~\label{app:models_arch}
\textbf{Tabular Foundation Models. } We use default parameters of such models.

TabPFN v2.6~\citep{hollmann_tabpfn_2023} is a transformer trained via prior data fitted networks to perform ICL over tabular data. Its encoder consists of 24 layers and 3 heads. Input features are grouped into blocks before embedding. A distinctive feature of TabPFN v2.6 is the use of 64 \emph{thinking rows}. These rows participate in attention and allow the transformer to perform intermediate reasoning over the provided context examples, but they do not correspond to real data records. The model supports a maximum context of 50.000 context samples.

TabICL~\citep{qu_tabicl_2025} uses a two-stage architecture: a column-interaction encoder (3 layers, 4 heads) that compresses each row into a fixed-dimensional embedding, followed by an ICL transformer (12 layers, 4 heads) that performs in-context learning over the resulting row embeddings.
The model supports a maximum context of 100.000 samples.

TabDPT v1.1~\citep{ma2026tabdpt} is a transformer-based ICL with 16 layers and 8 attention heads. It natively handles up to $100$ features; when a dataset exceeds this limit, Principal Component Analysis (PCA) is applied to compress the feature space to $100$ dimensions before inference. 

\textbf{Classical machine learning models. }
All models are trained with Optuna hyperparameter search (30 trials, 3-fold CV).
Random Forest starts from $n\_{estimators}=200$ trees with unconstrained depth, Optuna searches over $n\_{estimators}\in \{50,500\}$, $max\_{depth}\in\{3,5,8,10,15,20,None\}$,
${min\_samples\_split}\in\{2,20\}$ and $min\_samples\_leaf\in\{1, \max(10,\lfloor n/200\rfloor)\}$.

For LightGBM, 
Optuna searches over $n\_estimators\in\{50,500\}$, $\eta\in\{0.01,0.3\}$,
$max\_depth\in\{3,12\}$, $num\_leaves\in\{15,\min(255,\max(31,3p))\}$
(where $p$ is the number of features),
${subsample}\in\{0.5,1\}$,
and $\ell_1,\ell_2$ regularisation $\in\{10^{-4},10\}$.

TabNet starts with $n_d=n_a=32$, $n\_steps=3$, $\gamma=1.3$,
batch size $256$, virtual batch size $64$, up to $200$ epochs with patience $15$.
Optuna searches over $n_d\in\{8,16,32,64\}$, $n\_{steps}\in[3,8]$,
$\gamma\in[1,2]$, and $\eta\in[10^{-4},10^{-1}]$.

MLP is a three-layer network $(256, 256, 128)$, each followed by batch normalisation and ReLU activation. Optuna tunes the learning rate $\eta \in \{10^{-4}, 10^{-1}\}$, weight decay
$\in [10^{-6}, 10^{-2}]$, batch size $\in \{128, 256, 512\}$, and optimiser $\in \{\text{SGD}, \text{Adam}, \text{AdamW}\}$.
Training runs for up to $100$ epochs with early stopping after $15$ epochs.

\section{Hardware and computation resources}~\label{app:hardware}
All experiments were conducted on Ubuntu 24.04 LTS. The machine is equipped with dual Intel Xeon Silver 4416+ CPU, with 30 physical cores and 503 GiB of system memory. 
For accelerated computing, the system includes two NVIDIA L40S GPUs, each with approximately 46 GB of VRAM, giving roughly 90 GB of total GPU memory across the node. The experiments were run with CUDA 13.0. This setup was used for model inference and attack execution.

\section{Performance of the target models}~\label{app:targets_performance}
Table~\ref{tab:train-test-accuracy-roc-datasets} shows how well the models can make predictions. The results indicate overfitting, particularly wrt Locations dataset. While most models achieve near-perfect training accuracy, test accuracy is lower, with gaps of up to 0.26. This suggests that models are more likely to memorise training patterns than to generalise. Tabnet and RF demonstrate the strongest overfitting on the Locations dataset, while TabICL exhibits the smallest gap and the best test performance, suggesting superior generalisation. Overfitting is generally weaker on Dropout Success, but it is still notable for TabDPT, which has almost perfect training accuracy yet much lower test accuracy. This suggests limited generalisation despite a strong fit to the training set.

\begin{table}[ht!]
\centering
\scriptsize
\include{models_accuracy}
\caption{Average of the target model train and test accuracies (acc) for Location and Dropout Success datasets.}
\label{tab:train-test-accuracy-roc-datasets}
\end{table}

\section{Attack performance on six datasets}~\label{app:attack_performance}
\begin{figure}[ht!]
\centerline{\includegraphics[width=0.9\columnwidth]{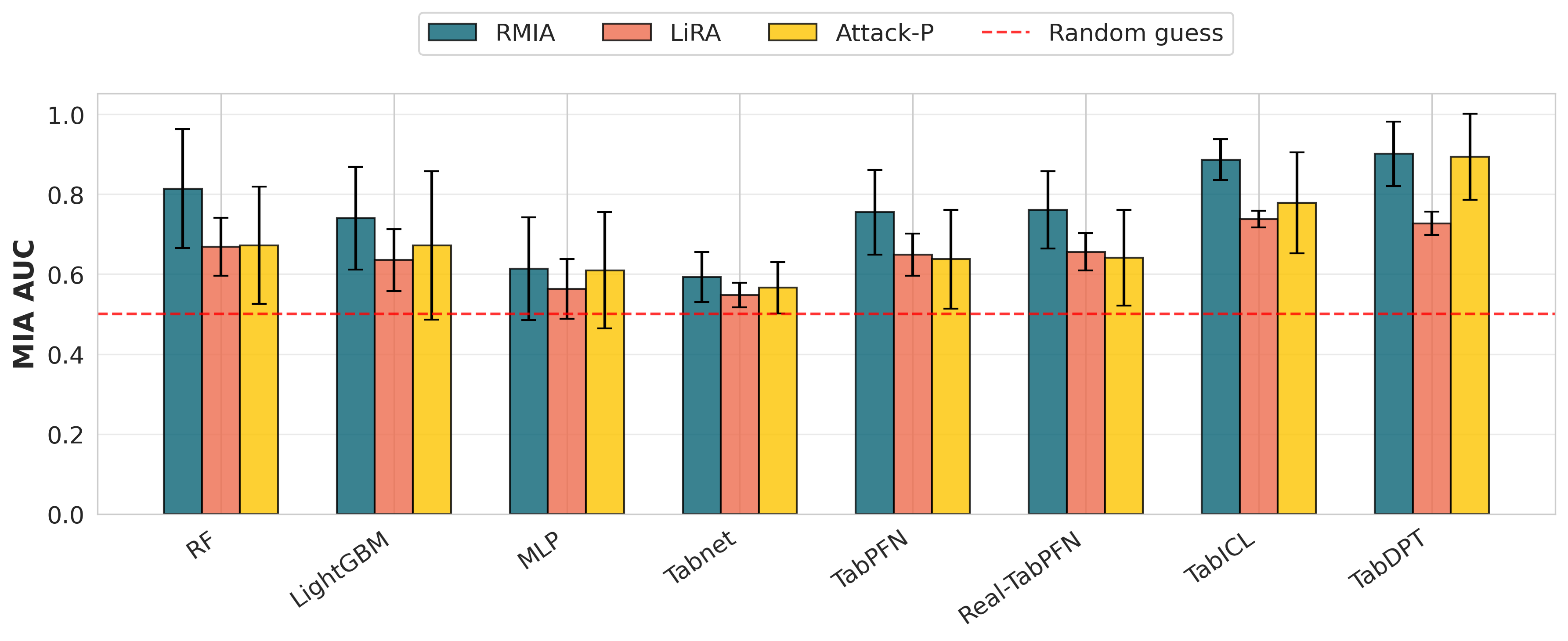}}
\caption{Comparison of MIAs across classical machine learning and tabular foundation models. Bars show mean MIA AUC obtained from six datasets, with error bars indicating variation across datasets.}
\label{fig:mia_performance}
\end{figure}

Figure~\ref{fig:mia_performance} shows that RMIA is generally the strongest attack across six datasets. Furthermore, tabular FMs, especially TabDPT and TabICL, exhibit higher membership leakage than most classical machine learning models.

\section{Attack performance over layers}~\label{app:layers_performance}
\begin{figure}[ht!]
    \centering
    
    \begin{subfigure}[b]{\columnwidth}
        \centering
        \includegraphics[width=0.8\textwidth]{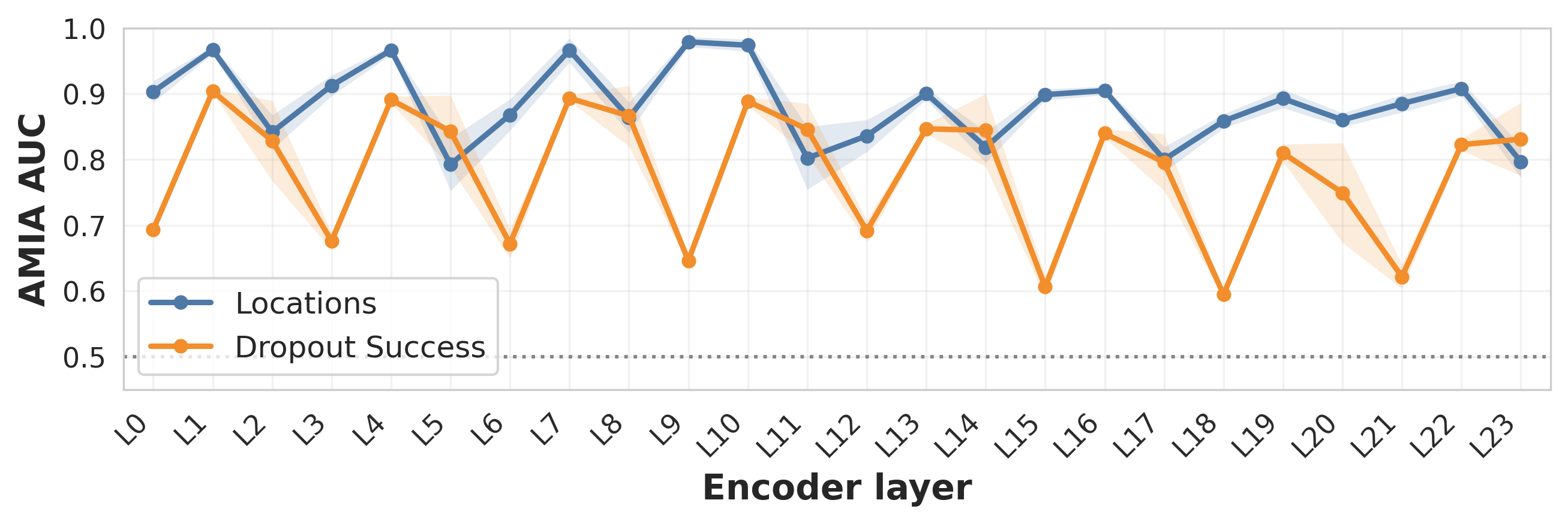}
    \end{subfigure}
    \vfill
    \begin{subfigure}[b]{\columnwidth}
        \centering
        \includegraphics[width=0.8\textwidth]{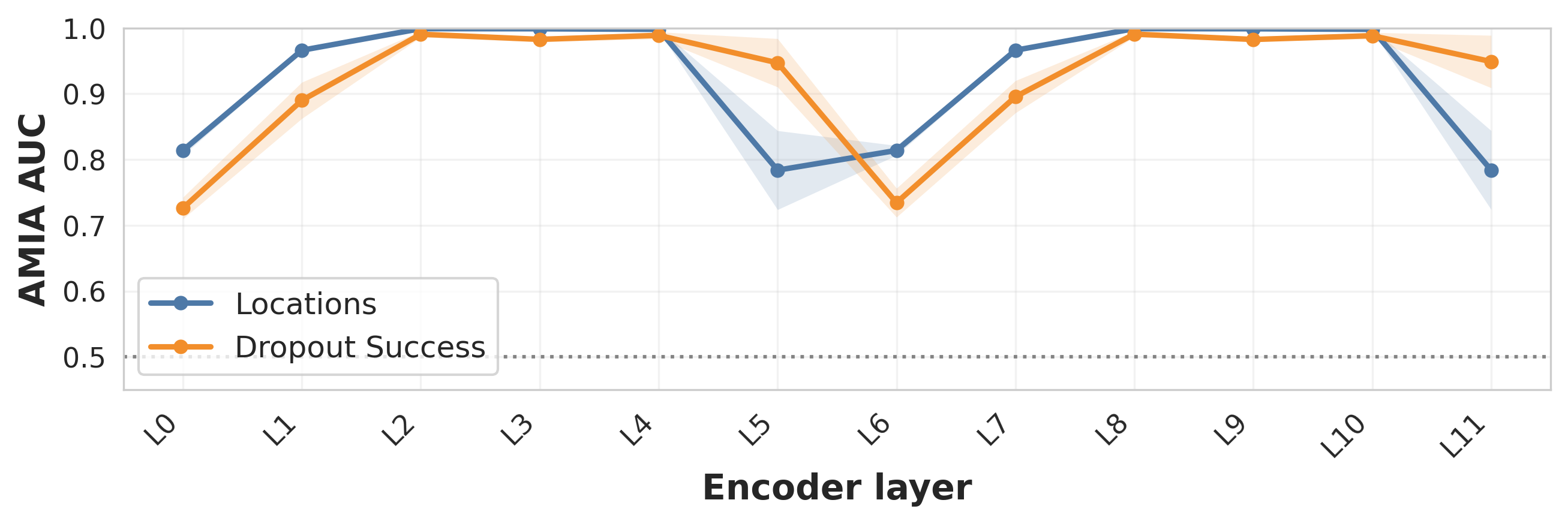}
    \end{subfigure}
    \vfill
    \begin{subfigure}[b]{\columnwidth}
        \centering
        \includegraphics[width=0.8\textwidth]{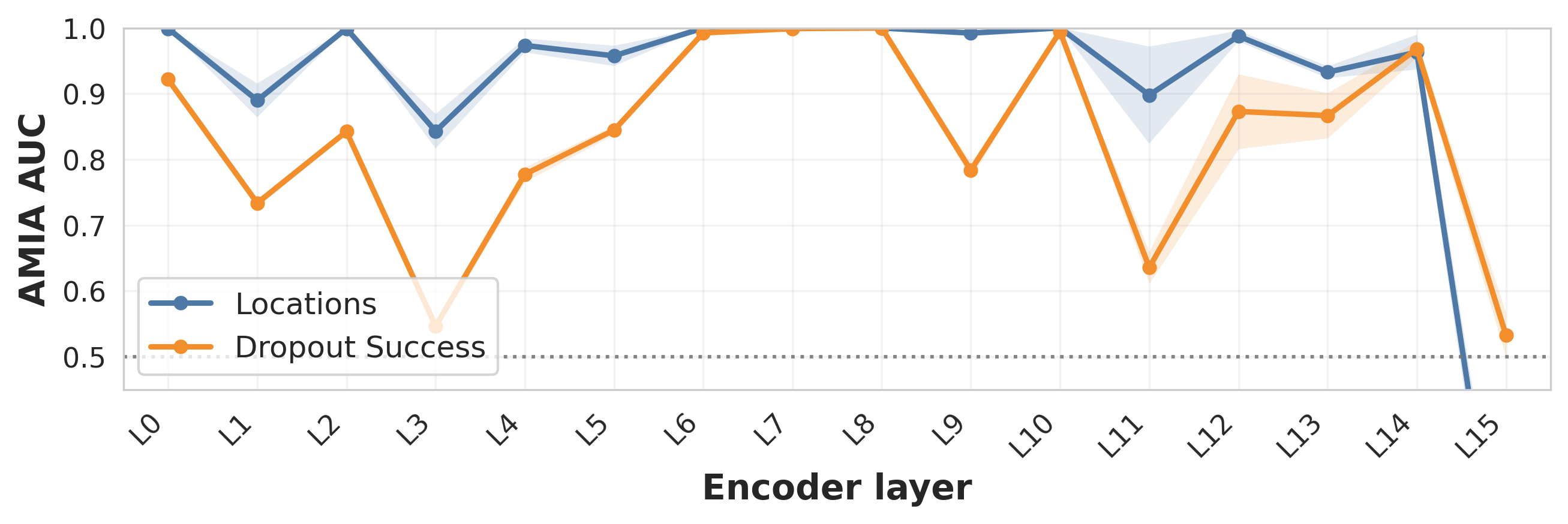}
    \end{subfigure}
    
    \caption{Per-layer AMIA AUC for TabPFN, TabICL and TabDPT respectively.}
    \label{fig:attack_layers}
\end{figure}

Figure~\ref{fig:attack_layers} shows that the attention-based membership signal varies substantially across layers, with several layers reaching near-perfect separability. 
The final AMIA ROC AUC is high because the aggregate score is supported by many highly discriminative layers. 
Low AUC layers contribute noise or an inverted signal, but the majority of layers provide a strong positive membership signal, so the aggregate AMIA score remains highly discriminative.

\section{AMIA performance on six datasets}~\label{app:amia_performance}
\begin{figure}[ht!]
\centerline{\includegraphics[width=0.8\columnwidth]{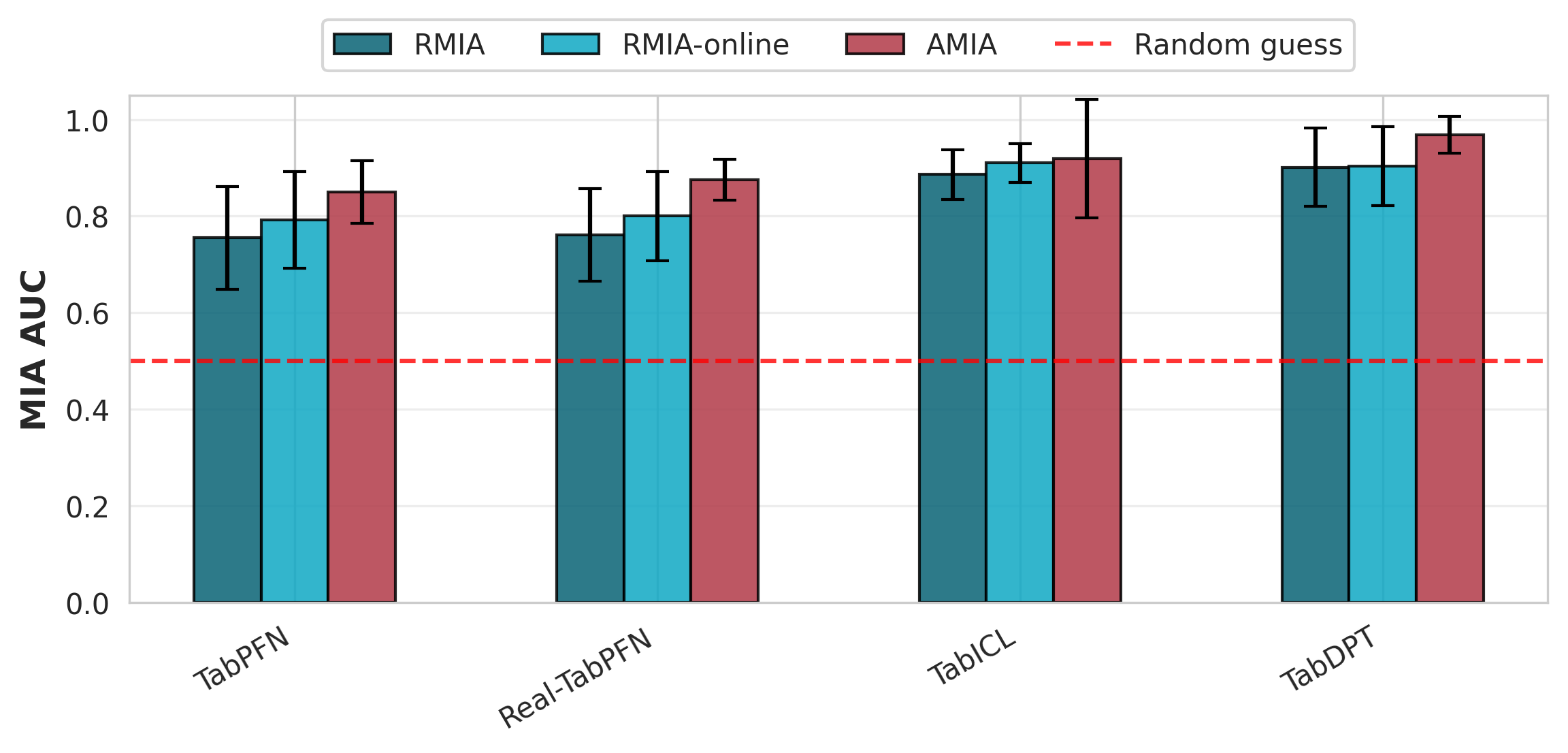}}
\caption{Comparison of AMIA with offline and online RMIA in tabular foundation models across six datasets.}
\label{fig:off_on_performance}
\end{figure}

Figure~\ref{fig:off_on_performance} provides the comparison of AMIA with both versions of RMIA. While offline uses trained reference models on randomly sampled datasets, avoiding any training on test queries, online trains reference models separately for each target data (test query $x$), in which IN models are trained containing $x$ in their training set.

AMIA achieves the highest average AUC for all models, especially TabDPT, while RMIA-online improves over offline RMIA for TabPFN, Real-TabPFN, and TabICL.

\section{Computational costs}~\label{app:computational_costs}
\begin{figure}[ht!]
\centerline{\includegraphics[width=0.9\columnwidth]{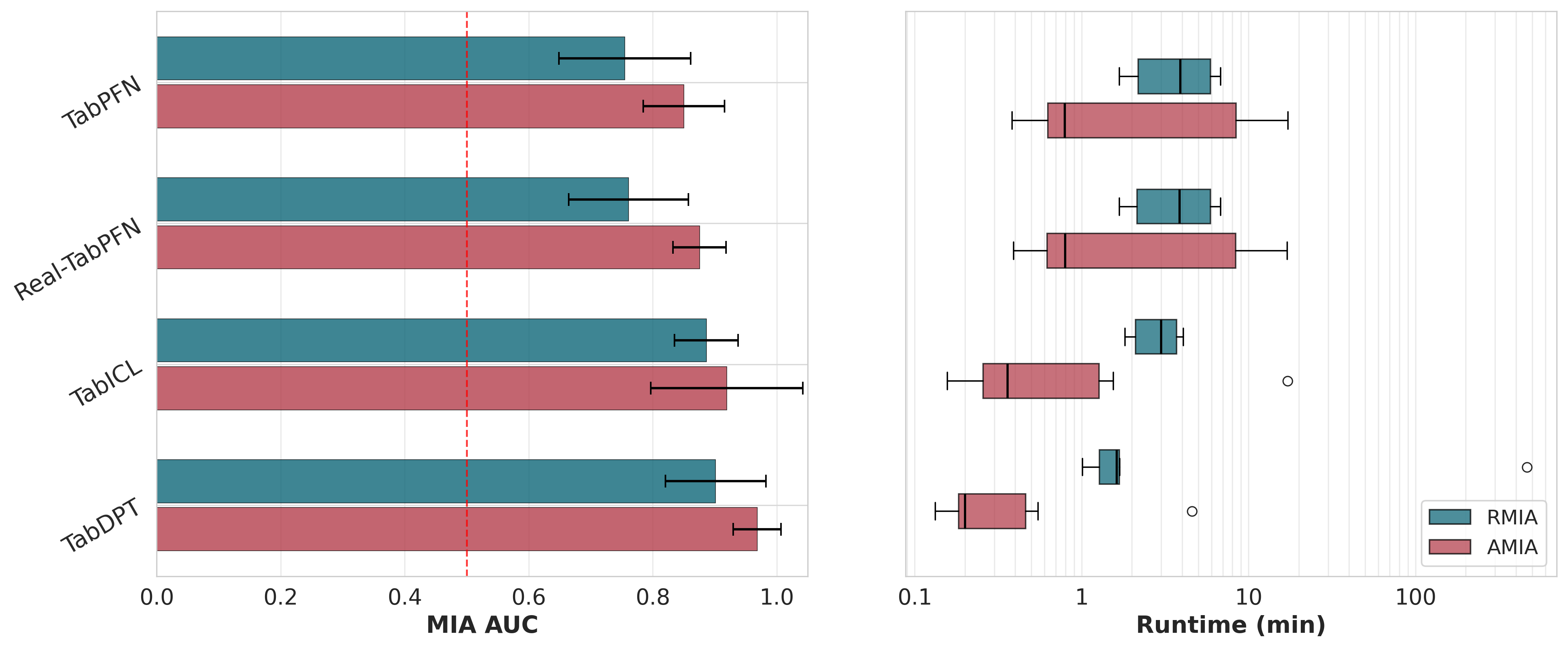}}
\caption{Attack effectiveness and runtime for AMIA and RMIA on tabular foundation models. The left panel reports mean MIA AUC across datasets; the right panel shows the distribution of per-dataset attack runtimes.}
\label{fig:runtime}
\end{figure}

For a clear comparison between attack strength and attack cost, Figure~\ref{fig:runtime} reports the execution time on the right panel. AMIA generally gives higher AUC, while also providing runtime advantage.

AMIA is consistently more computationally efficient than RMIA across all evaluated models. For TabDPT, the median runtime of AMIA is approximately 12 seconds, whereas RMIA requires around 49 seconds. A similar behaviour is observed for TabICL, where AMIA completes within seconds while RMIA requires several minutes.

The difference becomes evident for TabPFN and Real-TabPFN, where RMIA incurs particularly high computational costs. Importantly, the reported RMIA runtimes assume the use of two GPUs, whereas AMIA operates using only a single GPU. Therefore, the actual computational overhead of RMIA is substantially greater.

\section{Context size}~\label{app:context_size}
\begin{figure*}[ht]
\centerline{\includegraphics[width=0.8\textwidth]{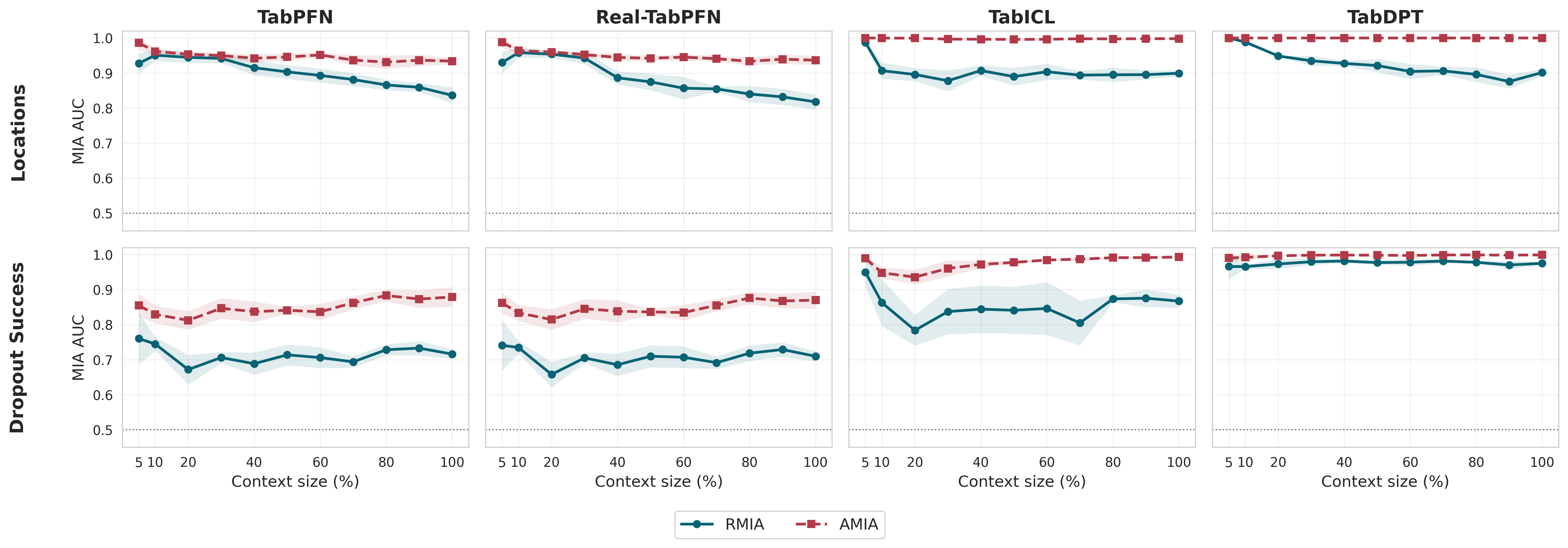}}
\caption{Evolution of membership inference attack risk as the context size increases. Context sizes corresponds to subsets from 5\% to 100\% of the context set pool. Points show mean AUC across seeds and shades indicate standard deviation.}
\label{fig:context_size}
\end{figure*}

Figure~\ref{fig:context_size} illustrates the evolution of AMIA and RMIA at different context sizes. This means that $D_{ctx}$ is progressively subsampled, from a small fraction of the available context rows to the complete context set. The audit pool is kept fixed across context sizes.

AMIA remains consistently strong over all context sizes, showing that leakage can persist even when the exposed context is reduced. However, RMIA is affected because the model's prediction scores change with context size. This divergence suggests that increasing $D_{ctx}$ can dilute membership information in the output probabilities used by RMIA, while simultaneously amplifying attention-based memorisation signals. With more context rows available, the model can distribute predictive evidence throughout the context, reducing the marginal effect of any single member on the output score. In contrast, AMIA directly observes whether the query focuses attention on specific context rows, a phenomenon that becomes more pronounced the larger the set of candidates from which the model can select.

\section{Defences ablation study}~\label{app:defence_ablation}
\begin{figure}[ht!]
\centerline{\includegraphics[width=0.9\columnwidth]{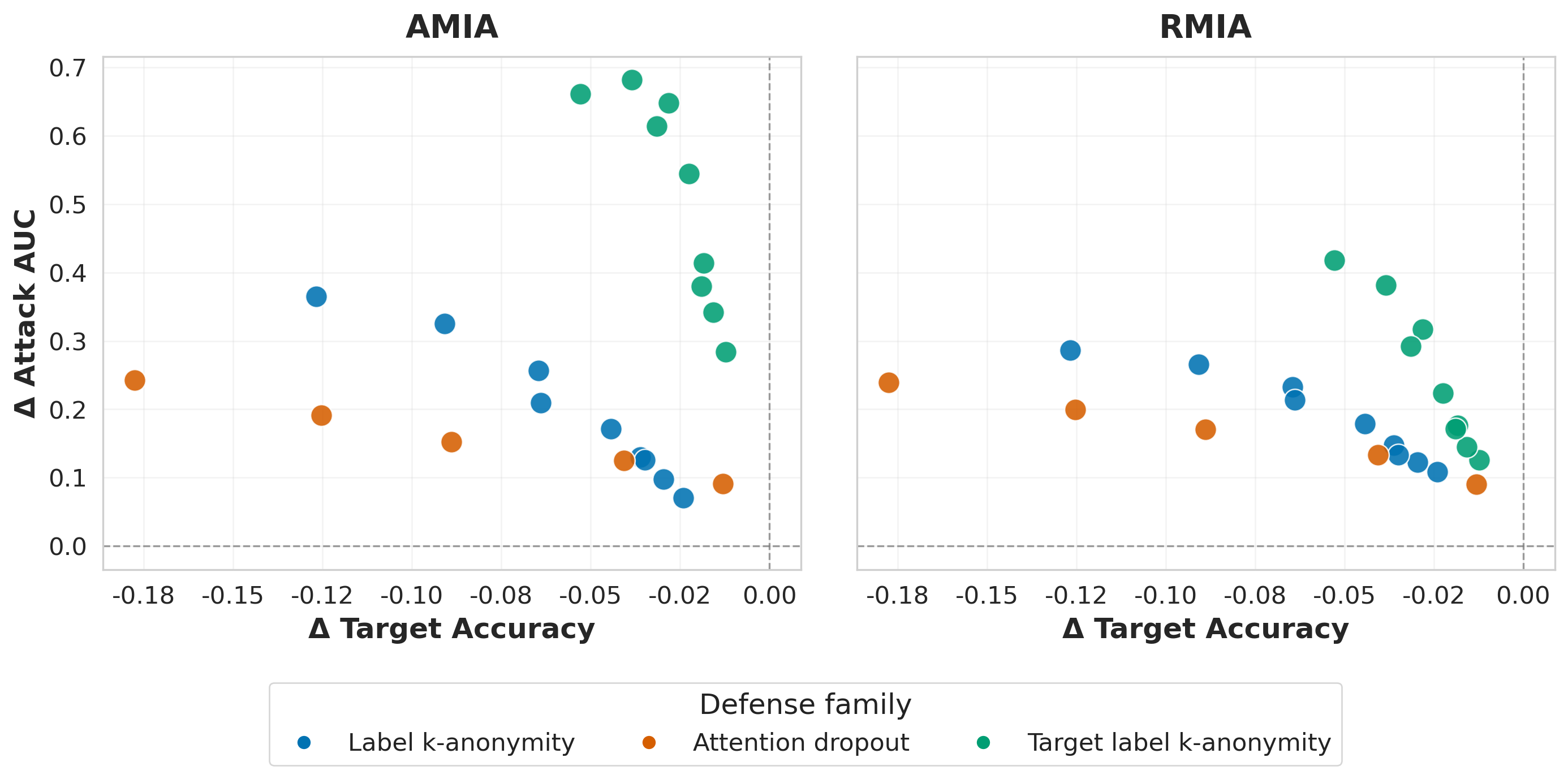}}
\caption{Privacy-utility effect of defence variants. Each point is one defence configuration, averaged across models. Points in superior right quadrant provide stronger privacy gains with higher model utility preservation.}
\label{fig:defences_ablation}
\end{figure}

To evaluate the privacy-utility tradeoff of the proposed target label $k$-anonymity defence, we compare it against two other proposed defences in Figure~\ref{fig:defences_ablation}. 

The first strategy, label $k$-anonymity, replaces each context row's key vector with the centroid of its $k$ same-label rows in key space, varying $k \in \{3, 5, 10\}$ and the soft interpolation coefficient $\alpha \in \{0, 0.3, 0.5\}$. Instead of transforming the high-risk queries, it uniformly transforms all queries regardless of their risk score. The second, attention dropout, regularises the row-attention weights by randomly zeroing out attention entries with probability $p \in \{0.1, 0.3, 0.5\}$. Besides, we automatically select the top layers which are defined as the layers with the highest baseline per-layer AMIA AUC. The defence first identifies which layers leak membership most strongly, then applies attention dropout only to those layers rather than all layers. This avoids changing all the layers and induce in higher predictive performance lost. Larger $p$ means stronger randomisation, usually more privacy protection but also more risk of reducing accuracy.

For each family and configuration, we report the reduction in AMIA and RMIA AUC relative to the undefended model alongside the corresponding accuracy drop, forming a privacy-utility trade-off curve across models and datasets.

In general, attention dropout is the least efficient among all defences. The member's key uniqueness in score space is still present, but the mean signal across layers is systematically reduced by the dropout probability. 
The AMIA score across queries is therefore a mixture of the full signal and noise. Thus, the attack can still rank members above non-members because the residual structure in the non-zeroed weights preserves the original ordering. Label $k$-anonymity, by contrast, consistently produce higher privacy gain because it directly targets the key uniqueness that drives concentration. However, this  approach also results in high predictive performance costs.
Target label $k$-anonymity dominates among the three defences for both AMIA and RMIA in terms of privacy and utility.

\section{Effect of fine-tuning}~\label{app:finetuning}
\begin{figure}[ht!]
\centerline{\includegraphics[width=0.75\columnwidth]{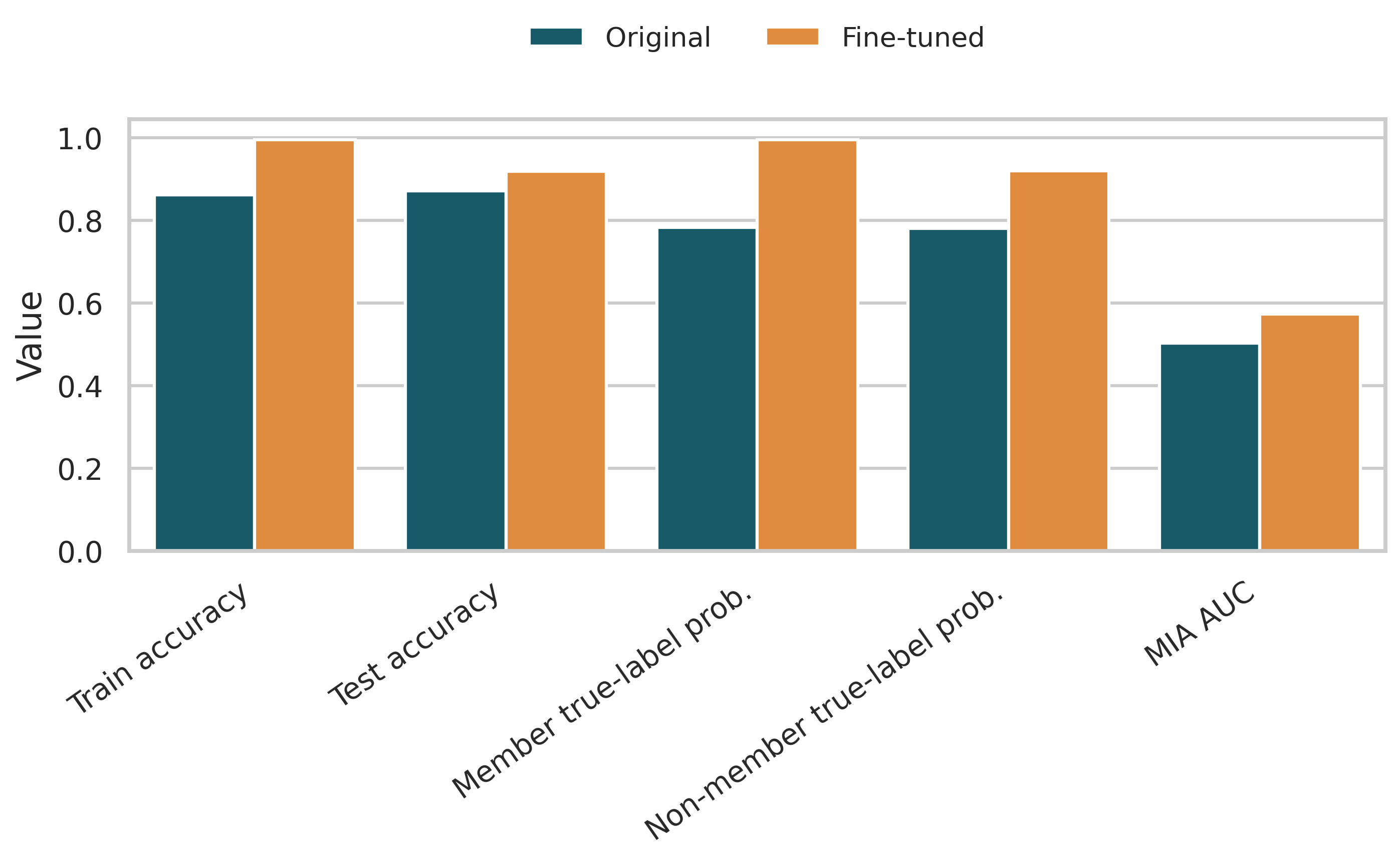}}
\caption{Fine-tuning utility and RMIA effectiveness.}
\label{fig:ft_train}
\end{figure}

Figure~\ref{fig:ft_train} compares the original and fine-tuned TabDPT models across accuracy, true label confidence, and MIA AUC. Such results are obtained with Purchases10 dataset. Fine-tuning improves task performance and increases true label confidence, especially on fine-tuning members. This utility gain is accompanied by increased MIA vulnerability.

\end{document}

%% file: 2datasets-rmia-lira-attackp.tex
\resizebox{\columnwidth}{!}{%
\renewcommand{\arraystretch}{1.15}
\begin{tabular}{llccc}
\toprule
\textbf{Dataset} & \textbf{Model} & \textbf{RMIA} & \textbf{LiRA} & \textbf{Attack-P} \\
\midrule
\multirow{8}{*}{Locations}
 & RF          & $\textbf{0.968} \pm 0.025$ & $0.739 \pm 0.016$ & $0.782 \pm 0.067$ \\
 & LightGBM    & $\textbf{0.854} \pm 0.042$ & $0.713 \pm 0.023$ & $0.841 \pm 0.039$ \\
 & TabNet      & $\textbf{0.694} \pm 0.027$ & $0.593 \pm 0.021$ & $0.683 \pm 0.026$ \\
 & MLP         & $0.820 \pm 0.012$ & $0.698 \pm 0.011$ & $\textbf{0.878} \pm 0.023$ \\ \cdashline{2-5}
 & TabPFN      & $\textbf{0.836} \pm 0.023$ & $0.703 \pm 0.017$ & $0.712 \pm 0.009$ \\
 & Real-TabPFN & $\textbf{0.818} \pm 0.021$ & $0.697 \pm 0.017$ & $0.700 \pm 0.008$ \\
 & TabICL      & $\textbf{0.899} \pm 0.008$ & $0.742 \pm 0.015$ & $0.814 \pm 0.016$ \\
 & TabDPT      & $0.902 \pm 0.010$ & $0.731 \pm 0.016$ & $\textbf{0.964} \pm 0.022$ \\
\midrule
\multirow{8}{*}{Dropout Success}
 & RF          & $\textbf{0.855} \pm 0.066$ & $0.685 \pm 0.024$ & $0.618 \pm 0.048$ \\
 & LightGBM    & $\textbf{0.746} \pm 0.034$ & $0.627 \pm 0.015$ & $0.583 \pm 0.017$ \\
 & TabNet      & $\textbf{0.636} \pm 0.059$ & $0.577 \pm 0.035$ & $0.594 \pm 0.047$ \\
 & MLP         & $\textbf{0.588} \pm 0.037$ & $0.545 \pm 0.016$ & $0.560 \pm 0.021$ \\ \cdashline{2-5}
 & TabPFN      & $\textbf{0.716} \pm 0.013$ & $0.633 \pm 0.007$ & $0.555 \pm 0.013$ \\
 & Real-TabPFN & $\textbf{0.710} \pm 0.015$ & $0.630 \pm 0.008$ & $0.554 \pm 0.014$ \\
 & TabICL      & $\textbf{0.867} \pm 0.019$ & $0.719 \pm 0.012$ & $0.653 \pm 0.027$ \\
 & TabDPT      & $\textbf{0.975} \pm 0.005$ & $0.753 \pm 0.016$ & $0.911 \pm 0.029$ \\
\bottomrule
\end{tabular}
}

%% file: finetuned.tex
\begin{tabular}{lcc}
\toprule
\textbf{Model} & \textbf{RMIA} & \textbf{AMIA} \\
\midrule
Original    & 0.867 & \textbf{1.000} \\
Fine-tuned  & 0.540 & \textbf{0.997} \\
\bottomrule
\end{tabular}

%% file: datasets.tex
\renewcommand{\arraystretch}{1.1}
\begin{tabular}{lccc}
\toprule
\textbf{Dataset} & $\textbf{\textit{n} samples}$ & $\textbf{\textit{d} features}$ & $\textbf{\textit{c} classes}$  \\
\midrule
MIC~\cite{ds:mic}                 & 1,699  & 111 & 8  \\
Locations~\cite{ds:location-unprocessed}           & 1,711  & 446 & 10 \\
Credit Rating~\cite{simonetto2024tabularbench}       & 2,029  & 30  & 10 \\
Dropout Success~\cite{ds:dropout_success}     & 4,424  & 36  & 3  \\
Financial Indicators US Stocks~\cite{ds:stocks} & 4,392  & 230 & 2  \\
URL~\cite{simonetto2024tabularbench}                 & 11,430 & 82  & 2  \\
Purchases10~\cite{ds:purchases-unprocessed}         & 19,763 & 599 & 10 \\
\bottomrule
\end{tabular}

%% file: models_accuracy.tex
\resizebox{\columnwidth}{!}{%
\renewcommand{\arraystretch}{1.15}
\begin{tabular}{lcccc}
\toprule
 & \multicolumn{2}{c}{\textbf{Locations}} & \multicolumn{2}{c}{\textbf{Dropout Success}} \\
\cmidrule(lr){2-3} \cmidrule(lr){4-5}
\textbf{Model} & \textbf{Train acc.} & \textbf{Test acc.} & \textbf{Train acc.} & \textbf{Test acc.} \\
\midrule
RF & 0.989 $\pm$ 0.009 & 0.760 $\pm$ 0.018 & 0.924 $\pm$ 0.044 & 0.766 $\pm$ 0.010 \\
LightGBM & 1.000 $\pm$ 0.000 & 0.812 $\pm$ 0.015 & 0.922 $\pm$ 0.027 & 0.768 $\pm$ 0.007 \\
MLP & 1.000 $\pm$ 0.000 & 0.818 $\pm$ 0.007 & 0.787 $\pm$ 0.050 & 0.685 $\pm$ 0.051 \\
Tabnet & 0.951 $\pm$ 0.067 & 0.694 $\pm$ 0.042 & 0.882 $\pm$ 0.073 & 0.727 $\pm$ 0.011 \\
\cdashline{1-5}
TabPFN & 0.996 $\pm$ 0.003 & 0.829 $\pm$ 0.018 & 0.862 $\pm$ 0.012 & \textbf{0.782} $\pm$ 0.005 \\
Real-TabPFN & 0.993 $\pm$ 0.005 & 0.816 $\pm$ 0.024 & 0.861 $\pm$ 0.014 & 0.781 $\pm$ 0.004 \\
TabICL & 0.999 $\pm$ 0.001 & \textbf{0.871} $\pm$ 0.013 & 0.943 $\pm$ 0.014 & 0.780 $\pm$ 0.006 \\
TabDPT & 1.000 $\pm$ 0.000 & 0.801 $\pm$ 0.011 & 0.996 $\pm$ 0.002 & 0.773 $\pm$ 0.008 \\
\bottomrule
\end{tabular}
}

%% file: refs.bib
@article{garg2022can,
  title={What can transformers learn in-context? a case study of simple function classes},
  author={Garg, Shivam and Tsipras, Dimitris and Liang, Percy S and Valiant, Gregory},
  journal={Advances in neural information processing systems},
  volume={35},
  pages={30583--30598},
  year={2022}
}

@article{reddy2023mechanistic,
  title={The mechanistic basis of data dependence and abrupt learning in an in-context classification task},
  author={Reddy, Gautam},
  journal={arXiv preprint arXiv:2312.03002},
  year={2023}
}

@article{brown2020language,
  title={Language models are few-shot learners},
  author={Brown, Tom and Mann, Benjamin and Ryder, Nick and Subbiah, Melanie and Kaplan, Jared D and Dhariwal, Prafulla and Neelakantan, Arvind and Shyam, Pranav and Sastry, Girish and Askell, Amanda and others},
  journal={Advances in neural information processing systems},
  volume={33},
  pages={1877--1901},
  year={2020}
}

@article{han2025understanding,
  title={Understanding emergent in-context learning from a kernel regression perspective},
  author={Han, Chi and Wang, Ziqi and Zhao, Han and Ji, Heng},
  journal={Transactions on Machine Learning Research},
  year={2025}
}

@misc{qu_tabicl_2025,
	title = {{TabICL}: {A} {Tabular} {Foundation} {Model} for {In}-{Context} {Learning} on {Large} {Data}},
	shorttitle = {{TabICL}},
	doi = {10.48550/arXiv.2502.05564},
	language = {en},
	urldate = {2025-05-06},
	publisher = {arXiv},
	author = {Qu, Jingang and Holzmüller, David and Varoquaux, Gaël and Morvan, Marine Le},
	month = feb,
	year = {2025},
	note = {arXiv:2502.05564 [cs]},
}

@misc{hollmann_tabpfn_2023,
	title = {{TabPFN}: {A} {Transformer} {That} {Solves} {Small} {Tabular} {Classification} {Problems} in a {Second}},
	shorttitle = {{TabPFN}},
	doi = {10.48550/arXiv.2207.01848},
	language = {en},
	urldate = {2025-05-05},
	publisher = {arXiv},
	author = {Hollmann, Noah and Müller, Samuel and Eggensperger, Katharina and Hutter, Frank},
	month = sep,
	year = {2023},
	note = {arXiv:2207.01848 [cs]},
}

@article{hollmann_tabpfn_2025,
	title = {{TabPFN}: {Accurate} predictions on small data with a tabular foundation model},
	volume = {637},
	issn = {0028-0836, 1476-4687},
	doi = {10.1038/s41586-024-08328-6},
	language = {en},
	number = {8045},
	urldate = {2025-05-05},
	journal = {Nature},
	author = {Hollmann, Noah and Müller, Samuel and Purucker, Lennart and Krishnakumar, Arjun and Körfer, Max and Hoo, Shi Bin and Schirrmeister, Robin Tibor and Hutter, Frank},
	month = jan,
	year = {2025},
	pages = {319--326},
}

@article{d2024context,
  title={In-context learning for extreme multi-label classification},
  author={D'Oosterlinck, Karel and Khattab, Omar and Remy, Fran{\c{c}}ois and Demeester, Thomas and Develder, Chris and Potts, Christopher},
  journal={arXiv preprint arXiv:2401.12178},
  year={2024}
}

@article{duan2024privacy,
  title={On the privacy risk of in-context learning},
  author={Duan, Haonan and Dziedzic, Adam and Yaghini, Mohammad and Papernot, Nicolas and Boenisch, Franziska},
  journal={arXiv preprint arXiv:2411.10512},
  year={2024}
}

@inproceedings{carey2024dp,
  title={Dp-tabicl: In-context learning with differentially private tabular data},
  author={Carey, Alycia N and Bhaila, Karuna and Edemacu, Kennedy and Wu, Xintao},
  booktitle={2024 IEEE International Conference on Big Data (BigData)},
  pages={1552--1557},
  year={2024},
  organization={IEEE}
}

@article{zaree2026attenmia,
  title={AttenMIA: LLM Membership Inference Attack through Attention Signals},
  author={Zaree, Pedram and Mamun, Md Abdullah Al and Dong, Yue and Alouani, Ihsen and Abu-Ghazaleh, Nael},
  journal={arXiv preprint arXiv:2601.18110},
  year={2026}
}

@article{erickson2026tabarena,
  title={Tabarena: A living benchmark for machine learning on tabular data},
  author={Erickson, Nick and Purucker, Lennart and Tschalzev, Andrej and Holzm{\"u}ller, David and Desai, Prateek and Salinas, David and Hutter, Frank},
  journal={Advances in Neural Information Processing Systems},
  volume={38},
  year={2026}
}

@inproceedings{arik2021tabnet,
  title={Tabnet: Attentive interpretable tabular learning},
  author={Arik, Sercan {\"O} and Pfister, Tomas},
  booktitle={Proceedings of the AAAI conference on artificial intelligence},
  volume={35},
  number={8},
  pages={6679--6687},
  year={2021}
}

@article{ma2026tabdpt,
  title={TabDPT: Scaling Tabular Foundation Models on Real Data},
  author={Ma, Junwei and Thomas, Valentin and Hosseinzadeh, Rasa and Labach, Alex and Cresswell, Jesse and Golestan, Keyvan and Yu, Guangwei and Caterini, Anthony L and Volkovs, Maks},
  journal={Advances in Neural Information Processing Systems},
  volume={38},
  pages={172692--172722},
  year={2026}
}

@article{arazi2026tabstar,
  title={TabSTAR: A Tabular Foundation Model for Tabular Data with Text Fields},
  author={Arazi, Alan and Shapira, Eilam and Reichart, Roi},
  journal={Advances in Neural Information Processing Systems},
  volume={38},
  pages={172108--172161},
  year={2026}
}

@article{zhang2026mitra,
  title={Mitra: Mixed synthetic priors for enhancing tabular foundation models},
  author={Zhang, Xiyuan and Maddix Robinson, Danielle and Yin, Junming and Erickson, Nick and Ansari, Abdul Fatir and Han, Boran and Zhang, Shuai and Akoglu, Leman and Faloutsos, Christos and Mahoney, Michael and others},
  journal={Advances in neural information processing systems},
  volume={38},
  pages={15795--15840},
  year={2026}
}

@article{zeng2025tabflex,
  title={Tabflex: Scaling tabular learning to millions with linear attention},
  author={Zeng, Yuchen and Dinh, Tuan and Kang, Wonjun and Mueller, Andreas C},
  journal={arXiv preprint arXiv:2506.05584},
  year={2025}
}

@article{hu2022membership,
  title={Membership inference attacks on machine learning: A survey},
  author={Hu, Hongsheng and Salcic, Zoran and Sun, Lichao and Dobbie, Gillian and Yu, Philip S and Zhang, Xuyun},
  journal={ACM Computing Surveys (CSUR)},
  volume={54},
  number={11s},
  pages={1--37},
  year={2022},
  publisher={ACM New York, NY}
}

@article{hu2023defenses,
  title={Defenses to membership inference attacks: A survey},
  author={Hu, Li and Yan, Anli and Yan, Hongyang and Li, Jin and Huang, Teng and Zhang, Yingying and Dong, Changyu and Yang, Chunsheng},
  journal={ACM Computing Surveys},
  volume={56},
  number={4},
  pages={1--34},
  year={2023},
  publisher={ACM New York, NY}
}

@article{byun2025risk,
  title={Risk In Context: Benchmarking Privacy Leakage of Foundation Models in Synthetic Tabular Data Generation},
  author={Byun, Jessup and Lin, Xiaofeng and Ward, Joshua and Cheng, Guang},
  journal={arXiv preprint arXiv:2507.17066},
  year={2025}
}

@article{carvalho2023survey,
  title={Survey on privacy-preserving techniques for microdata publication},
  author={Carvalho, T{\^a}nia and Moniz, Nuno and Faria, Pedro and Antunes, Lu{\'\i}s},
  journal={ACM Computing Surveys},
  volume={55},
  number={14s},
  pages={1--42},
  year={2023},
  publisher={ACM New York, NY}
}

@article{ward2025tables,
  title={When Tables Leak: Attacking String Memorization in LLM-Based Tabular Data Generation},
  author={Ward, Joshua and Gu, Bochao and Wang, Chi-Hua and Cheng, Guang},
  journal={arXiv preprint arXiv:2512.08875},
  year={2025}
}

@inproceedings{shokri2017membership,
  title={Membership inference attacks against machine learning models},
  author={Shokri, Reza and Stronati, Marco and Song, Congzheng and Shmatikov, Vitaly},
  booktitle={2017 IEEE symposium on security and privacy (SP)},
  pages={3--18},
  year={2017},
  organization={IEEE}
}

@inproceedings{ye2022enhanced,
  title={Enhanced membership inference attacks against machine learning models},
  author={Ye, Jiayuan and Maddi, Aadyaa and Murakonda, Sasi Kumar and Bindschaedler, Vincent and Shokri, Reza},
  booktitle={Proceedings of the 2022 ACM SIGSAC conference on computer and communications security},
  pages={3093--3106},
  year={2022}
}

@article{zarifzadeh2023low,
  title={Low-cost high-power membership inference attacks},
  author={Zarifzadeh, Sajjad and Liu, Philippe and Shokri, Reza},
  journal={arXiv preprint arXiv:2312.03262},
  year={2023}
}

@inproceedings{carlini2022membership,
  title={Membership inference attacks from first principles},
  author={Carlini, Nicholas and Chien, Steve and Nasr, Milad and Song, Shuang and Terzis, Andreas and Tramer, Florian},
  booktitle={2022 IEEE symposium on security and privacy (SP)},
  pages={1897--1914},
  year={2022},
  organization={IEEE}
}

@inproceedings{wen_membership_2024,
    address = {Salt Lake City UT USA},
    title = {Membership {Inference} {Attacks} {Against} {In}-{Context} {Learning}},
    isbn = {979-8-4007-0636-3},
    doi = {10.1145/3658644.3690306},
    language = {en},
    urldate = {2026-05-19},
    booktitle = {Proceedings of the 2024 on {ACM} {SIGSAC} {Conference} on {Computer} and {Communications} {Security}},
    publisher = {ACM},
    author = {Wen, Rui and Li, Zheng and Backes, Michael and Zhang, Yang},
    month = dec,
    year = {2024},
    pages = {3481--3495},
}

@inproceedings{chang2025context,
  title={Context-aware membership inference attacks against pre-trained large language models},
  author={Chang, Hongyan and Shamsabadi, Ali Shahin and Katevas, Kleomenis and Haddadi, Hamed and Shokri, Reza},
  booktitle={Proceedings of the 2025 Conference on Empirical Methods in Natural Language Processing},
  pages={7299--7321},
  year={2025}
}

@article{ran2025lora,
  title={Lora-leak: Membership inference attacks against lora fine-tuned language models},
  author={Ran, Delong and He, Xinlei and Cong, Tianshuo and Wang, Anyu and Li, Qi and Wang, Xiaoyun},
  journal={arXiv preprint arXiv:2507.18302},
  year={2025}
}

@article{chen2023overconfidence,
  title={Overconfidence is a dangerous thing: Mitigating membership inference attacks by enforcing less confident prediction},
  author={Chen, Zitao and Pattabiraman, Karthik},
  journal={arXiv preprint arXiv:2307.01610},
  year={2023}
}

@article{choi2025contextleak,
  title={Contextleak: Auditing leakage in private in-context learning methods},
  author={Choi, Jacob and Cao, Shuying and Dong, Xingjian and Banayeeanzade, Amin and Zhu, Wang Bill and Jia, Robin and Karimireddy, Sai Praneeth},
  journal={arXiv preprint arXiv:2512.16059},
  year={2025}
}

@incollection{dwork2025differential,
  title={Differential privacy},
  author={Dwork, Cynthia},
  booktitle={Encyclopedia of Cryptography, Security and Privacy},
  pages={649--652},
  year={2025},
  publisher={Springer}
}

@inproceedings{bhusal2026privacy,
  title={Privacy Preserving In-Context-Learning Framework for Large Language Models},
  author={Bhusal, Bishnu and Acharya, Manoj and Kaur, Ramneet and Samplawski, Colin and Roy, Anirban and Cobb, Adam D and Chadha, Rohit and Jha, Susmit},
  booktitle={Proceedings of the AAAI Conference on Artificial Intelligence},
  volume={40},
  number={42},
  pages={35303--35312},
  year={2026}
}

@article{pera2026sok,
  title={SoK: Challenges in Tabular Membership Inference Attacks},
  author={P{\^e}ra, Cristina and Carvalho, T{\^a}nia and Cordy, Maxime and Antunes, Lu{\'\i}s},
  journal={arXiv preprint arXiv:2601.15874},
  year={2026}
}

@article{duan2023flocks,
  title={Flocks of stochastic parrots: Differentially private prompt learning for large language models},
  author={Duan, Haonan and Dziedzic, Adam and Papernot, Nicolas and Boenisch, Franziska},
  journal={Advances in Neural Information Processing Systems},
  volume={36},
  pages={76852--76871},
  year={2023}
}

@inproceedings{tang2024privacy,
  title={Privacy-preserving in-context learning with differentially private few-shot generation},
  author={Tang, Xinyu and Shin, Richard and Inan, Huseyin and Manoel, Andre and Mireshghallah, Niloofar and Lin, Zinan and Gopi, Sivakanth and Kulkarni, Janardhan and Sim, Robert},
  booktitle={International conference on learning representations},
  volume={2024},
  pages={33058--33077},
  year={2024}
}

@inproceedings{amin2024private,
  title={Private prediction for large-scale synthetic text generation},
  author={Amin, Kareem and Bie, Alex and Kong, Weiwei and Kurakin, Alexey and Ponomareva, Natalia and Syed, Umar and Terzis, Andreas and Vassilvitskii, Sergei},
  booktitle={Findings of the Association for Computational Linguistics: EMNLP 2024},
  pages={7244--7262},
  year={2024}
}

@inproceedings{wu2024privacy,
  title={Privacy-preserving in-context learning for large language models},
  author={Wu, Tong and Panda, Ashwinee and Wang, Jiachen Tianhao and Mittal, Prateek},
  booktitle={International Conference on Learning Representations},
  volume={2024},
  pages={20005--20040},
  year={2024}
}

@inproceedings{zheng2024locally,
  title={Locally differentially private in-context learning},
  author={Zheng, Chunyan and Sun, Keke and Zhao, Wenhao and Zhou, Haibo and Jiang, Lixing and Song, Shaoyang and Zhou, Chunlai},
  booktitle={Proceedings of the 2024 Joint International Conference on Computational Linguistics, Language Resources and Evaluation (LREC-COLING 2024)},
  pages={10686--10697},
  year={2024}
}

@article{Koskela2025DifferentiallyPI,
  title={Differentially Private In-Context Learning with Nearest Neighbor Search},
  author={Antti Koskela and Tejas D. Kulkarni and Laith Zumot},
  journal={ArXiv},
  year={2025},
  volume={abs/2511.04332}
}

@article{German2025TabMIAAB,
  title={Tab-MIA: A Benchmark Dataset for Membership Inference Attacks on Tabular Data in LLMs},
  author={Eyal German and Sagiv Antebi and Daniel Samira and Asaf Shabtai and Yuval Elovici},
  journal={ArXiv},
  year={2025},
  volume={abs/2507.17259}
}

@article{Sweeney2002kAnonymityAM,
  title={k-Anonymity: A Model for Protecting Privacy},
  author={Latanya Sweeney},
  journal={Int. J. Uncertain. Fuzziness Knowl. Based Syst.},
  year={2002},
  volume={10},
  pages={557-570}
}

@article{DomingoFerrer2002PracticalDM,
  title={Practical Data-Oriented Microaggregation for Statistical Disclosure Control},
  author={Josep Domingo-Ferrer and Josep Maria Mateo-Sanz},
  journal={IEEE Trans. Knowl. Data Eng.},
  year={2002},
  volume={14},
  pages={189-201}
}

@article{vaswani2017attention,
  title={Attention is all you need},
  author={Vaswani, Ashish and Shazeer, Noam and Parmar, Niki and Uszkoreit, Jakob and Jones, Llion and Gomez, Aidan N and Kaiser, {\L}ukasz and Polosukhin, Illia},
  journal={Advances in neural information processing systems},
  volume={30},
  year={2017}
}

@article{truex2019demystifying,
  title={Demystifying membership inference attacks in machine learning as a service},
  author={Truex, Stacey and Liu, Ling and Gursoy, Mehmet Emre and Yu, Lei and Wei, Wenqi},
  journal={IEEE transactions on services computing},
  volume={14},
  number={6},
  pages={2073--2089},
  year={2019},
  publisher={IEEE}
}

@misc{ds:mic,
  author       = {Lennart Purucker},
  title        = {MIC},
  howpublished = {OpenML},
  note = {id=46943; Accessed May 2026}
}

@misc{ds:stocks,
  author       = {Yayun Li},
  title        = {Financial Indicators US Stocks},
  howpublished = {OpenML},
  note = {id=46567; Accessed May 2026}
}

@misc{ds:dropout_success,
  author       = {Valentim Realinho and Mónica Vieira Martins and Jorge Machado and Luís Baptista},
  title        = {{Predict Students' Dropout and Academic Success}},
  year         = {2021},
  howpublished = {UCI Machine Learning Repository},
  note         = {Accessed Jan 2026}
}

@misc{ds:location-unprocessed,
  title={Foursquare Global-scale Check-in Dataset with User Social Networks},
  author={Dingqi YANG},
  note = {Accessed Jan 2026}
}

@misc{ds:purchases-unprocessed,
  title={Acquire Valued Shoppers Challenge},
  publisher={Kaggle},
  howpublished={Kaggle},
  author={Will Cukierski},
  note = {Accessed Jan 2026}
}

@article{simonetto2024tabularbench,
  title={Tabularbench: Benchmarking adversarial robustness for tabular deep learning in real-world use-cases},
  author={Simonetto, Thibault and Ghamizi, Salah and Cordy, Maxime},
  journal={Advances in Neural Information Processing Systems},
  volume={37},
  pages={78394--78430},
  year={2024}
}
